\documentclass[fleqn,usenatbib]{mnras}

\usepackage{newtxtext,newtxmath}
\usepackage{subcaption}
\usepackage{longtable}
\usepackage{float}

\usepackage[T1]{fontenc}

\DeclareRobustCommand{\VAN}[3]{#2}
\let\VANthebibliography\thebibliography
\def\thebibliography{\DeclareRobustCommand{\VAN}[3]{##3}\VANthebibliography}

\usepackage{graphicx}	
\usepackage{amsmath}	

\newcommand{\kms}{km s$^{-1}$}
\newcommand{\FeH}{[\mathrm{Fe/H}]}
\newcommand{\DTD}{\mathrm{DTD}}

\newcommand{\logDTD}{\log{\DTD}}

\newcommand{\NRRintProb}{23}

\def\neww#1{{\textcolor{black}{#1}}}
\def\newww#1{{\textcolor{black}{#1}}}

\title[Intermediate-age RRL in MC clusters]{RR Lyrae Stars in Intermediate-age Magellanic Clusters: Membership Probabilities and Delay Time Distribution}

\author[B. Cuevas-Otahola et al.]{
Bolivia Cuevas-Otahola,$^{1,2}$\thanks{E-mail: b.cuevas.otahola@gmail.com (BCO). The first two authors contributed equally.}
Cecilia Mateu,$^{3}$\thanks{E-mail: cmateu@fcien.edu.uy}
Ivan Cabrera-Ziri,$^{4,5}$
Gustavo Bruzual,$^{1}$
\newauthor
Fabiola Hernández-Pérez$^{6,7}$, Gladis Magris$^{8}$ and Holger Baumgardt$^{9}$
\\
$^{1}$Instituto de Radioastronom\'ia y Astrof\'isica, Universidad Nacional Aut\'onoma de M\'exico, Morelia, Michoac\'an, 58089, M\'exico\\
$^{2}$Departamento de Matem\'aticas, FCE, Benem\'erita Universidad Aut\'onoma de Puebla, Puebla, 72000, M\'exico\\
$^{3}$Departamento de Astronomía, Facultad de Ciencias, Universidad de la República, Iguá 4225, 14000, Montevideo, Uruguay\\
$^{4}$Astronomisches Rechen-Institut, Zentrum für Astronomie der Universität Heidelberg, Monchhofstraße 12-14, D-69120 Heidelberg, Germany\\
$^{5}$Vyoma GmbH, Karl-Theodor-Straße 55, 80803 Munich\\
$^{6}$Centro de Estudios de Física del Cosmos de Aragón (CEFCA), Plaza San Juan 1, 44001 Teruel, Spain.\\
$^{7}$Universidad Internacional de Valencia (VIU), C/Pintor Sorolla 21, 46002, Valencia, Spain.\\
$^{8}$Centro de Investigaciones de Astronomía, CIDA, Mérida, Venezuela \\
$^{9}$School of Mathematics and Physics, The University of Queensland, St Lucia, QLD 4072, Australia
}

\date{Accepted 2025 June 30. Received 2025 June 29; in original form 2024 November 19}

\pubyear{2015}

\begin{document}
\label{firstpage}
\pagerange{\pageref{firstpage}--\pageref{lastpage}}
\maketitle

\begin{abstract}
Recent works have challenged our canonical view of RR Lyrae (RRL) stars as tracers of exclusively old populations ($\gtrsim10$~Gyr) by proposing a fraction of these stars to be of intermediate ages ($\sim$2-5~Gyr). Since it is currently not possible to infer stellar ages directly for individual RRL stars, our goal in this work is to search for these in association to intermediate-age clusters whose reliable ages can then be safely be attributed to the RRL. We used the Gaia DR3 Specific Object Study and OGLE IV public catalogues to search for RRL stars around stellar clusters older than 1~Gyr in the Large and Small Magellanic Clouds. 
Modelling membership probabilities based on proper motion and photometric distance we obtained a list of 259 RRL stars associated with Magellanic clusters. Of these, \NRRintProb\ RRL are likely members of 10 intermediate-age clusters: 3 and 7 in the Small and Large Magellanic Clouds, respectively. By modelling the inferred expectation values of the number of RRL stars per cluster, we inferred the delay time distribution of the RRL in three age ranges. \neww{For the old population ($>8$~Gyr) we \neww{find $2.5^{+0.4}_{-0.3}$ RRL$/10^5 M_\odot$. For the young (1-2 Gyr) and intermediate age (2-8 Gyr) populations we find rates of $0.34^{+0.17}_{-0.12}$ and $0.071^{+0.073}_{-0.041}$ RRL$/10^5 M_\odot$}, respectively, after further decontamination from control field tests}. While radial velocities are necessary for definitively confirming cluster memberships, the high-probability list of intermediate-age RRL stars presented here offers a promising opportunity for the first direct confirmation of these enigmatic stars. 

\end{abstract}

\begin{keywords}
keyword1 -- keyword2 -- keyword3
\end{keywords}



\section{Introduction}\label{Sec:intro}

RR Lyrae (RRL) stars, horizontal branch pulsators well known for their standard candle nature, have long been considered a quintessential Population II tracer. Abundant in Galactic globular clusters, where the variability class was first identified \citep{Bailey1902}, they have been longstanding tracers of the oldest Galactic stellar populations, 
mainly the halo \citep[e.g.][]{Saha1985,Wetterer1996,Vivas2006,Iorio2021,Medina2024}, thick disk \citep[e.g.][and refereces therein]{Layden1995a,Kinemuchi2006,Mateu2018} and bulge \citep[e.g.][]{Wegg2019,Kunder2022review}, and more recently --thanks to observational advances-- also in external galaxies \citep[see review by][]{Beaton2018}. Their popularity as tracers of the structure of the Milky Way (MW) has soared as deep multi-epoch surveys have published RRL catalogues covering increasingly larger areas \citep[e.g. QUEST-I, LONEOS, PanStarrs-1,][]{Vivas2004,Miceli2008,Sesar2017c}, eventually extending the entire sky \citep[e.g. ASAS-SN-II, Gaia Data Release 2 (DR2) and Data Release 3 (DR3), respectively)][]{Jayasinghe2019a,Clementini2018_DR2RRL,Clementini2023_SOS_GaiaDR3}. Their precise distances and relatively high luminosity, now combined with the all-sky kinematic information provided by the Gaia mission DR2  
\citep{GaiaCol_2018_DR2_survey}, have made them ideal to probe dynamics across large volumes of the Galaxy \citep[e.g.][]{Koposov2019_orphan,PriceWhelan2019,Ramos2020,Wegg2019,Kunder2022review,Iorio2021}.

Stellar evolution models for isolated stars predict RRL stars should only exist in stellar populations both old ($\gtrsim9-10$Gyr) and metal-poor enough ($\FeH <-0.5$) for the Horizontal Branch (HB) to cross the Instability Strip where the stars pulsate, becoming RRL \citep[e.g.][]{Smith1995,CatelanSmith2015}. This, combined with the observed abundance of RRL in predominantly or even exclusively old populations like Galactic globular clusters, ultra-faint dwarf galaxies, and galactic haloes, has contributed to their almost automatic association as tracers of exclusively old and metal-poor populations. For example, early studies of the star formation history of MW dwarf spheroidal (dSph) satellites specifically used the presence of RRL to establish that all dSph satellites contain an old and metal-poor population, regardless of how extended their star formation history had been \citep[e.g.][]{Grebel1998,Grebel2000b}, a claim later confirmed as photometry for the faintest main-sequence turn-offs has become accessible, validating the ubiquitous presence of an old and metal-poor component in MW dSph galaxies \citep[see review by][]{Tolstoy2009_review,Weisz2014}. 

This canonical view of RRL as tracers of old and metal-poor populations started meeting its first challenges from the metallicity perspective, as metal-rich RRL were first observed in the MW disc. \citet{Layden1995a,Layden1995b} first observed RRL with cold, thin disc kinematics which were later confirmed with spectroscopic data to be more metal-rich than their halo or thick disc counterparts. Similarly, \citet{Kunder2008} observed RRL at near-solar metallicities in the bulge and, more recently, \citet{Mullen2021} also report bona fide metal-rich RRL in their comprehensive catalogue, albeit at a small fraction.
These findings have recently been confirmed for solar neighbourhood RRL with new kinematic data from Gaia DR2 by \citet{Prudil2020}  and with new spectroscopic metallicity and $\alpha$-abundance measurements by \citet{Zinn2020} and \citet{Marsakov2019}, the latter proposing the thin disc kinematics imply a younger age for these RRL stars. From a theoretical standpoint, \citet{Bono1997a,Bono1997b} explored stellar evolution models with enhanced He abundances, leading to increased mass loss rates to be able to produce metal-rich RRL. These RRL are produced by more (initially) massive progenitors, at ages of $\sim$1~Gyr, much younger than typical RRL ages. Although the mass loss rates required for this mechanism to produce RRL may be at odds with the mass loss rates inferred from RGB stars \citep{Bobrick2022}, these models provide valuable insight into the pulsational properties that might be expected should
these objects form, e.g. via an alternative, more individualized evolutionary route, such as binary evolution.

More recently, two independent studies have found evidence that points at the existence of intermediate-age RRLs, challenging the canonical view, now from the age perspective. \citet[][hereafter S21]{Sarbadhicary_2021} inferred the Delay time Distribution (DTD) of RRL stars by jointly analysing the spatial distribution of OGLE-IV RRLs \citep{Soszynski2016} and the map of the Star Formation History of the Large Magellanic Cloud (LMC) from \citet{HarrisZaritsky2009}. The DTD quantifies the production rate (by unit initial mass) of a given type of object —in this case RRLs— as a function of time after a star formation burst. S21 found that as many as half (51\%) the LMC RRL stars have delay times (i.e. ages) between 1 and 8 Gyrs. In a completely separate study, \citet[][hereafter I21]{Iorio2021}  studied the kinematics  of Galactic RRL stars using astrometric data for the Gaia DR2 SOS RRL catalogue. They found a component of fast  rotating RRL stars whose kinematics are consistent with a disc population of an intermediate age of a few Gyrs. The rotation curve at $R>5$~kpc of $\approx230$\kms\ found for this component rotates much faster ($>40$\kms) than what would be expected for thick disc RRL. Their kinematics are also much colder, with a surprisingly low velocity dispersion of $\approx 20$\kms,  resembling that of intermediate populations of $\sim2$~Gyr, and about half the observed velocity dispersion ($\approx40$\kms) of older disc populations (10~Gyr) associated with canonical RRL stars \citep{SandersDas2018}. Similarly to S21, I21 propose there must be a population of much younger RRL in the disc. I21 and S21 thus reach a similar conclusion as S21 that intermediate-age RRL stars exist, based on completely independent data, methodologies, and even in two different galaxies (the MW and LMC) at different metallicity ranges.

These findings are surprising at first, as neither can be explained by canonical stellar evolution models. Both S21 and I21 point towards binary evolution as the culprit, providing a possible formation mechanism for these stars. This scenario had already been investigated by \citet{Karczmarek2017} who used the stellar synthesis code STARTRACK to explore the formation of binary evolution pulsators, i.e. RRL and other pulsators produced at unusual ages in binary systems with mass transfer events. More recently, \citet{Bobrick2022} used the MESA stellar synthesis code together with a model of Galactic stellar populations with a similar purpose. They both find RRL can be produced at intermediate ages by mass transfer in binary systems with an MS companion as a plausible formation channel.

The evidence provided by these works for the existence of intermediate-age RRLs remains indirect, however, since ages are not (and currently cannot) be estimated for individual RRL. 
A smoking gun to confirm that intermediate-age RRL exist indeed would be either to find RRL stars in binary systems, e.g. with a (main sequence) companion that can be dated, or to find them unequivocally associated to a simple stellar population with an intermediate age. The first \neww{strategy} was attempted by S21 using Gaia DR2, with no good candidates found. \neww{Previous} searches for RRL in binaries had resulted in very few candidate systems \citep[$\sim20$][]{Richmond2011,Liska2016,Kervella2019a,Kervella2019b}, with no robust identifications. \neww{ \citet{Prudil2019} and \citet{Hajdu2021} used the light-travel time effect to identify, respectively, 16 and 61  candidates for RRLs in binary systems in the direction of the Galactic Bulge. \citet{Prudil2019} report two candidates, OGLE-BLG-RRLYR-08185 and 13477, with companion stars likely to be more massive than the RRL component, while \citet{Hajdu2021} inferred (present-day) masses for the companions obtaining a trimodal distribution with peaks at $\sim0.6, \sim0.2$ and $\sim0.067~\mathrm{M}_\odot$. In both cases, since the variability signal observed in the $O-C$ curves can have explanations other than the light-travel time effect \citep{Skarka2018},  spectroscopic follow-up is needed to confirm the binary and, in that case, the nature and properties of the companion.
To the best of our knowledge,  currently the only two  RRLs \neww{confidently} known to be part of binary systems are OGLE-BLG-RRLYR-02792 \citep{Pietrzynski2012} and TU UMa \citep{Szeidl1986,Liska2016_TUUMa}, with the first being a binary evolution pulsator.} Neither are companions to young or intermediate-age stars. 
In this work, we attempt the second strategy\neww{, i.e., searching for RRLs associated to intermediate-age clusters}. 

The best chance to find any RRLs associated to intermediate-age clusters would seem to be to search for them in clusters sufficiently massive to be expected to host a non-zero number of RRL stars. S21 inferred the production rate of RRLs to be 1 in $10^5 M_{\sun}$ at intermediate-ages, about an order of magnitude lower than the mean rate of $\sim1$ in $10^4 M_{\sun}$ observed in Galactic globular clusters. This means that only relatively massive clusters ($>10^5 M_{\sun}$) would be expected to host at least one RRL. Such a rate may also explain why these RRL stars may have remained under the radar:  the MW famously hosts almost no intermediate age clusters, let alone massive ones \citep{Harriscat1996}. Massive clusters in the MW are exclusively old ($\gtrsim10$ Gyr); with young and intermediate-age clusters being so much less massive ($<\mathrm{few}\times10^3M_{\sun}$) they would not be expected to host even a single RRL star, according to S21’s results. 

The nearest system to host any bona fide intermediate-age clusters with masses of at least $10^5 M_{\sun}$ are the Magellanic Clouds. M31 hosts a population of even more massive clusters than the MW overall, with several exceeding $10^6M_{\sun}$. However, age determinations for intermediate-age cluster candidates in M31 \citep[][Cabrera-Ziri et al. in prep.]{Caldwell2011}, have been shown to be unreliable as they are derived from integrated spectra and the hot emission from horizontal branch stars can lead an old population to be mistaken for a much younger one unless accounted for \citep{Worthey1994,deFreitasPacheco1995,Beasley2002}. The ideal search place are then the Large and Small Magellanic Clouds (LMC and SMC, hereafter).

Our goal in this work is, then, to use the Gaia DR3 \citep{Clementini2023_SOS_GaiaDR3} and OGLE-IV \citep{Soszynski2016} RRL catalogues to identify stars with a high probability of being associated to intermediate-age clusters in the Magellanic Clouds and to use the results for the ensemble of stellar clusters to infer a rate at which intermediate age populations may produce RRL stars.  
In Section~\ref{Sec:Data} we describe the catalogue of RRL stars and of Magellanic clusters used in the search. In Sec.~\ref{Sec:model} we describe the inference model used to assign membership probabilities and to estimate the expected number of RRL members in each cluster. Because the expected rate of production of these stars in any individual cluster is expected to be low, we use the data for the ensemble of clusters to obtain an estimate of the production rate of RRLs per unit mass in a given age range, as we describe in Section~\ref{Sec:dtd_inference_model}. We discuss our results in Section~\ref{Sec:discussion} and summarise our conclusions in Section~\ref{Sec:conclusions}.

\section{Data}\label{Sec:Data}

We used LMC and SMC clusters data from the database of Structural Parameters of Local Group Star Clusters\footnote{\label{footnote:CatHolger}\url{https://people.smp.uq.edu.au/HolgerBaumgardt/globular/lgclusters/parameter.html}}, complemented with data from the literature as summarised in Table~\ref{Tab:clusters}. 
The table contains celestial coordinates, ages, masses, metallicities, proper motions, and line-of-sight distances to the clusters. The table also includes the \emph{initial} mass for each cluster ($\log M_i$), which will be necessary for the computation of the DTD. Initial masses were computed using main sequence lifetimes obtained from the PARSEC\footnote{Web interface at \url{http://stev.oapd.inaf.it/cgi-bin/cmd_3.7}} stellar evolutionary tracks \citep{PARSEC_Bressan2012,PARSEC_Chen2014,PARSEC_Chen2015,PARSEC_Tang2014} for the metallicity nearest to the cluster's, interpolating in age, and assuming  Initial Mass Function (IMF) parameters, lower and upper mass limits for the broken power-law IMF determined by \citet{Baumgardt2023} from a set of 120 MW and massive LMC/SMC clusters. We assume their corresponding errors to be the same as for the present-day mass $\log M$, reported in the table. This estimate ignores any mass the cluster may have lost due to dynamical effects.

\begin{figure*}
\begin{subfigure}{0.47\textwidth}
    \centering  \includegraphics[width=\textwidth]{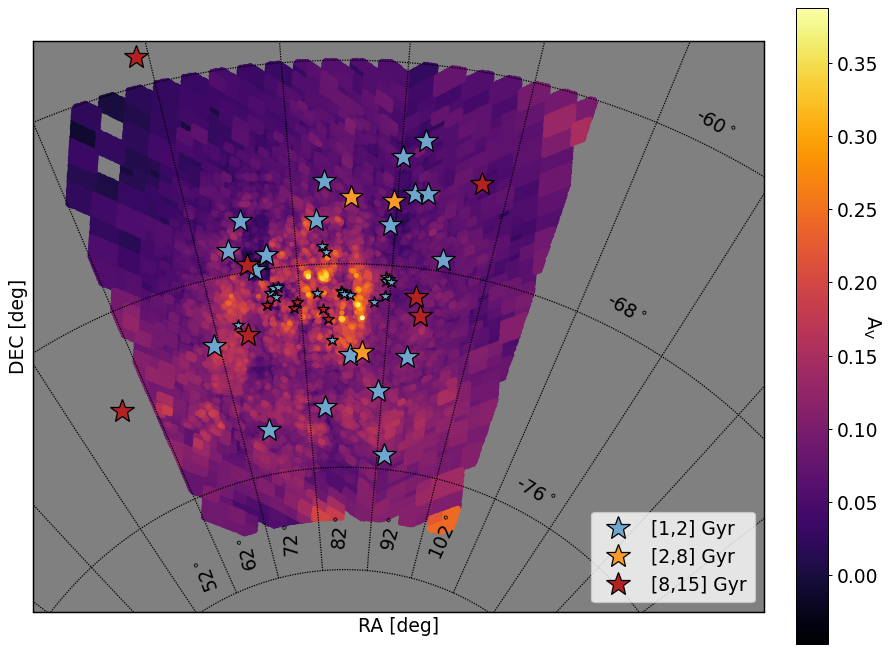}
    \caption{LMC}
\end{subfigure}
\begin{subfigure}{0.43\textwidth}
    \centering \includegraphics[width=\textwidth]{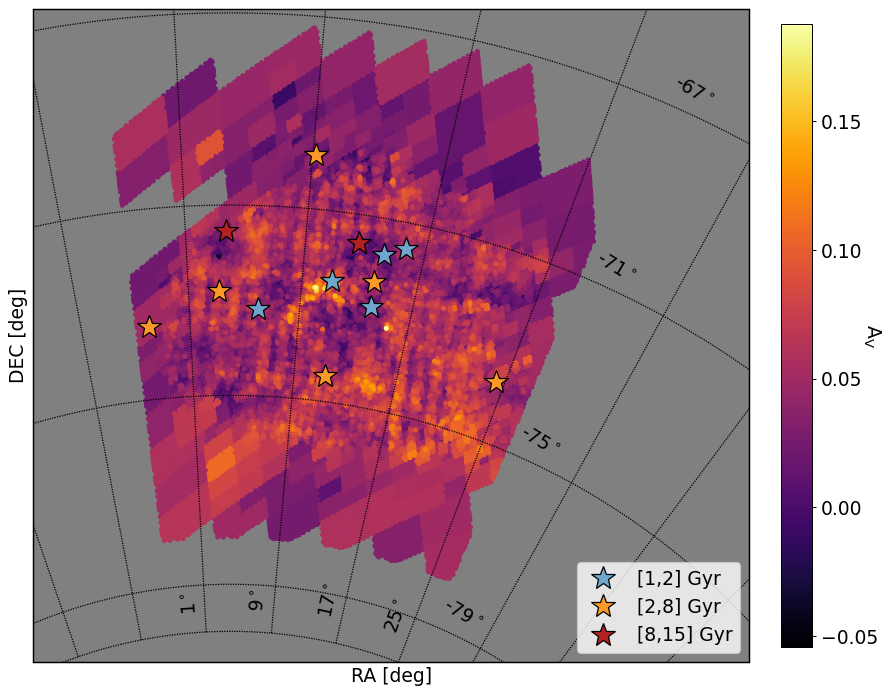}
    \caption{SMC}
\end{subfigure}
\caption{Spatial distribution of the stellar clusters in the LMC (left) and SMC (right) listed in Tab. \ref{Tab:clusters}. The colour scale shows the extinction map from \citet{Chen2022}. Star clusters with smaller symbol sizes were not considered for the analysis.}
\label{fig:ext_map}
\end{figure*}

\begin{figure*}
    \centering
    \includegraphics[width=2\columnwidth]{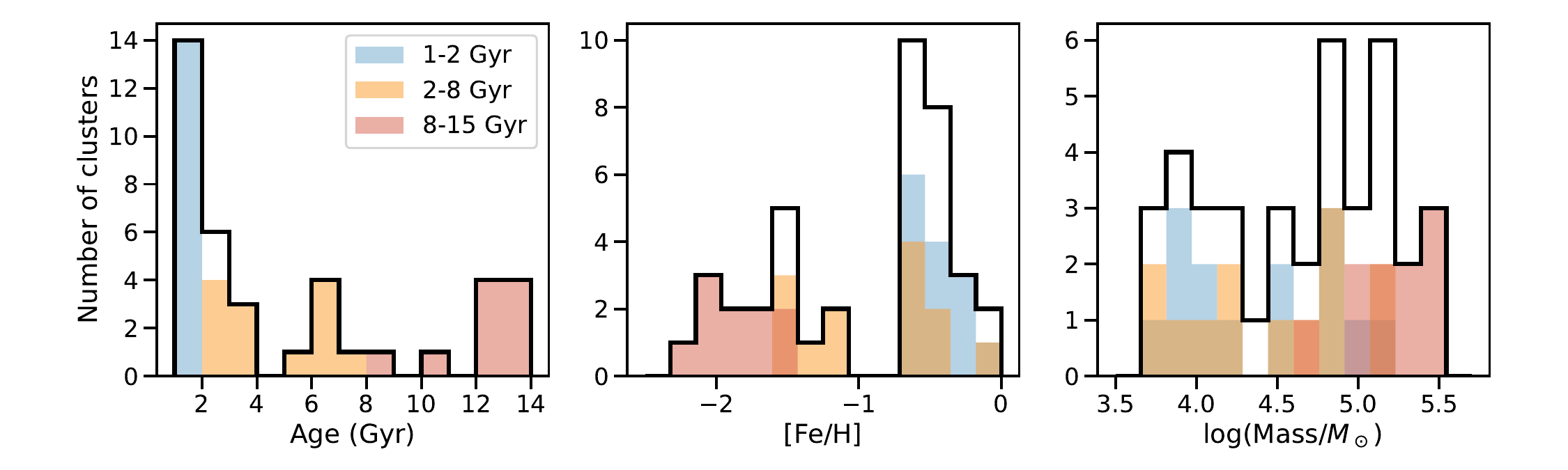}
    \caption{Distributions of age (\emph{left}), metallicity (\emph{centre}) and logMass (\emph{right}) for the LMC and SMC clusters from Table~\ref{Tab:clusters} used in this work in three age bins: young (1-2 Gyr), intermediate (2-8 Gyr) and old (>8 Gyr). }
    \label{fig:clusters_AMZ_hist.pdf}
\end{figure*}

Following S21, we have excluded --from the table and the analysis-- clusters in the bar region in the centre of the LMC to avoid crowding issues and high extinction regions in the disc prone to be also highly spatially variable, as this hinders a reliable estimation of the photometric distance to the RRL stars. The spatial distribution of the clusters in the LMC and SMC is shown in Figure~\ref{fig:ext_map}, respectively in the left and right panels. The extinction map from \citet{Chen2022} is shown by the colour scale. Clusters are shown with different symbols in the figure, according to three ages ranges: young (1-2~Gyr), intermediate (2-8~Gyr) and old (8-15 Gyr), each of which contains 16, 14 and 9 clusters in total, respectively. For these, as well as for the rest of the analysis, we have considered NGC~361 (age 8.1 Gyr) to belong to the intermediate-age bin since it is right on the edge and there are no other border line clusters like it at ages in the 8-10 Gyr range. The age, metallicity and mass distributions for the clusters are shown in Figure~\ref{fig:clusters_AMZ_hist.pdf}.

The RRL catalogue used in this work is the combination of the two main public surveys of RRL stars covering the Magellanic Clouds: OGLE-IV \citep{Soszynski2016} and the Specific Objects Study \citep[SOS, ][]{Clementini2023_SOS_GaiaDR3} catalogue provided with the Gaia Data Release 3 (DR3). The OGLE-IV catalogue is comprised by 39,082 and 6,369 RRL stars in the LMC and SMC, respectively. The Gaia DR3 SOS catalogue contains 271,779 RRL stars in total. The OGLE-IV catalogue was cross-matched to the Gaia DR3 \verb|gaia_source| table at a 1" tolerance and duplicates were removed based on Gaia DR3 \verb|source_id|. 

Photometric distances to the RRL were computed by using the $M_G-\FeH$ relation from \citet{Muraveva2018}. We assume each cluster's metallicity and extinction reported in Table~\ref{Tab:clusters} for all the RRLs in an extended vicinity of each cluster ($20\times r_h$), and using the \verb|phot_mean_g_mag| from the Gaia DR3 \verb|gaia_source| table as the mean $G$-band magnitude. By using only $G$-band data to compute distances we avoid crowding issues that affect $BP$ and $RP$ photometry in high density fields, but to which $G$-band photometry is much less prone. \neww{
By assuming the cluster's metallicity we also avoid the requirement of $\phi_{31}$ in order to estimate the metallicity. This would have forced us either to discard the nearly half of Gaia DR3 the sample without well-enough sampled light curves to have reliably estimated this parameter, or to assume a metallicity only for a fraction of the catalogue.} 
\neww{The reason for assuming each cluster’s metallicity is that, in the absence of data, if the assumed metallicity is wrong the distance to the RRL will definitely also be wrong. The converse, however, is not true: if an RRL does not belong to the cluster, assuming the wrong metallicity will still produce an incorrect distance, but this is of almost no consequence for background stars as it will simply cause the background distance distribution to be distorted. Although there may be some distance-metallicity combinations that, for an incorrectly-assumed metallicity, may make background RRL lie at the cluster’s distance, since the apparent distance distribution of the background will be inferred by our model from the data itself, this contribution should already be accounted for \footnote{\neww{We have found very similar results using the cluster's metallicity only for stars without available photometric metallicities, which gives us confidence that our results are not significantly affected by this assumption.}}. }

Finally, the extinction correction was made using the $V$-band extinction reported in Table~\ref{Tab:clusters} for each cluster, converted to the $G$-band with the transformation reported by \citet{Ramos2020} in their Appendix A, Equation~A.1, assuming for all stars $BP-RP=0.7$,  the mean colour of halo RRL in Gaia DR3. This procedure results in distance errors around $6\%$, assuming an error of 0.1~dex for the cluster's metallicity.

\section{Membership probability model}\label{Sec:model}

First, for the i-th cluster, we build a probabilistic mixture model where each star has a probability of being a member of the cluster or of the background, based on its parallax and proper motions, similarly to \citet{delPino2022} and \citet{PriceWhelan2019}, and described in detail for a general case in \citet{ForemanMackey2014_blog}. In this model the likelihood $\mathcal{L}$ of a given star is represented by a weighted mixture of its likelihood of being part of the cluster, given by $\mathcal{L}_{cl}$, or of the background, given by $\mathcal{L}_{bg}$, with a mixture weight parameter $f$, as:

\begin{equation}
    \mathcal{L} = f\mathcal{L}_{cl} + (1-f)\mathcal{L}_{bg}  \label{eq:L_mixture_model}
\end{equation}

We represent the likelihood of the star being a cluster member $\mathcal{L}_{cl}$ as the product of two Gaussian distributions: one in distance and one in proper motions. In each, the Gaussian's mean is given by the cluster's mean distance or proper motion accordingly and each standard deviation is the sum in quadrature of the cluster's intrinsic width, its error and the star's observational error in the corresponding attribute.
By doing this we have effectively taken the cluster's parameters as free and, under the assumption of Gaussian uncertainties, marginalised over them analytically using the properties of gaussian convolution by summing the standard deviations in quadrature.
This way, the likelihood of the data $D_n=(\varpi,\mu_{\alpha},\mu_{\delta})_n$ for the $n$-th star is given by the following expression

\begin{equation}\label{Eq:L_cluster}
   \mathcal{L}_{cl}(D_n|\theta_{cl}) = 
   \mathcal{N}(\vec{\mu}_n|\vec{\mu}_{cl},\sigma_\mu)
   \mathcal{N}(\varpi_n|\varpi_{cl},\sigma_\varpi)
\end{equation}

For the distance, the angular width is taken as the half light radius reported in Table~\ref{Tab:clusters} and converted to a line of sight width by using the cluster's distance, also listed in Table~\ref{Tab:clusters}, summed in quadrature with a 5\% distance error for all clusters with measured distances, or a 10\% for those without (an LMC distance of 50.1~kpc was assumed for these). For the proper motions, we take as the cluster's standard deviation the \neww{errors in proper motions summed in quadrature with the} observed intrinsic velocity dispersion listed in Table~\ref{Tab:clusters}, converted to proper motion using the cluster's assumed distance.

The data $D_n$ for all non-RRL stars is taken from the \verb|gaia_source| table of the Gaia DR3 catalogue and, as recommended by \citet{Lindegren2018}, only stars with $\mathrm{RUWE}<1.4$ are kept, in order to ensure their astrometric data is reliable (this filter applies to all stars, including RRL). For the RRL, proper motion errors are also taken from Gaia DR3, while for the parallax, instead of using the trigonometric parallax measured by Gaia, we use the photometric parallax $\varpi$ (and its corresponding error) computed as the inverse of the photometric distance computed in Sec.~\ref{Sec:Data}, and propagate errors accordingly. This way we take advantage of the precision offered by the $M_G-\FeH$ relation, which yields distance --and also parallax-- errors $<6\%$ (see previous section), much lower than typical Gaia DR3 parallax errors for individual stars in the LMC/SMC, which for most stars exceed $100\%$. 
\neww{Since RRLs distance errors are well below 20\%, the approximation that parallax is the inverse of distance is valid \citep[see e.g.][]{BailerJones2018}.}

\begin{table*}
\caption{Star Clusters in the Magellanic Clouds}

\begin{center}
\addtolength{\tabcolsep}{-0.4em}
\begin{tabular}{|l|l|r|r|r|r|r|r|r|r|r|r|r|r|r|r|r|}
\hline
  \multicolumn{1}{|c|}{IDs} &
  \multicolumn{1}{c|}{Galaxy} &
  \multicolumn{1}{c|}{Age} &
  \multicolumn{1}{c|}{RA} &
  \multicolumn{1}{c|}{DEC} &
  \multicolumn{1}{c|}{$\mu_{\rm RA}$} &
  \multicolumn{1}{c|}{$\mu_{\rm DEC}$} &
  \multicolumn{1}{c|}{logM} &
  \multicolumn{1}{c|}{logM$_i$} &
  \multicolumn{1}{c|}{Rh} &
  \multicolumn{1}{c|}{D} &
  \multicolumn{1}{c|}{$\sigma$} &
  \multicolumn{1}{r|}{Source} &
  \multicolumn{1}{c|}{Av} &
  \multicolumn{1}{c|}{Av ref} &
  \multicolumn{1}{c|}{Fe/H}  &
  \multicolumn{1}{c|}{Fe/H ref}\\
  \multicolumn{1}{|c|}{} &
  \multicolumn{1}{c|}{} &
  \multicolumn{1}{c|}{(Gyr)} &
  \multicolumn{1}{c|}{(deg)} &
  \multicolumn{1}{c|}{(deg)} &
  \multicolumn{1}{c|}{(mas/yr)} &
  \multicolumn{1}{c|}{(mas/yr)} &
  \multicolumn{1}{c|}{($\rm M_\odot$)} &
  \multicolumn{1}{c|}{($\rm M_\odot$)} &
  \multicolumn{1}{c|}{(pc)} &
  \multicolumn{1}{c|}{(kpc)} &
  \multicolumn{1}{c|}{(km/s)} &
  \multicolumn{1}{c|}{} &
  \multicolumn{1}{c|}{(mag)} &
  \multicolumn{1}{c|}{} &
  \multicolumn{1}{c|}{(dex)}  &
  \multicolumn{1}{c|}{}\\
\multicolumn{1}{|c|}{(1)} &
  \multicolumn{1}{c|}{(2)} &
  \multicolumn{1}{c|}{(3)} &
  \multicolumn{1}{c|}{(4)} &
  \multicolumn{1}{c|}{(5)} &
  \multicolumn{1}{c|}{(6)} &
  \multicolumn{1}{c|}{(7)} &
  \multicolumn{1}{c|}{(8)} &
  \multicolumn{1}{c|}{(9)} &
  \multicolumn{1}{c|}{(10)} &
  \multicolumn{1}{c|}{(11)} &
  \multicolumn{1}{r|}{(12)}  &
  \multicolumn{1}{c|}{(13)} &
  \multicolumn{1}{c|}{(14)}  &
  \multicolumn{1}{c|}{(15)} &
  \multicolumn{1}{c|}{(16)} &
  \multicolumn{1}{c|}{(17)}\\
  \hline
   Hodge11 &  LMC & 13.0000 & 93.5917 & -69.8483 &      1.48$\pm$0.03 & 0.99$\pm$0.03    &  5.3 $\pm$ 0.11 &	 5.90	&  10.94 & 51.8 & 	3.64	& M & 0.14 $\pm$ 0.08 & M & -2.06 & M\\
   Hodge14 &  LMC &  1.8197 & 82.1625 & -73.6300 &      2.01$\pm$0.03 & 0.36$\pm$0.03    &  4.1 $\pm$ 0.18 &	 4.47	&   9.66 & 51.3 & 	1.12	& M & 0.33 $\pm$ 0.15 & M & -0.66 & M\\
    Hodge4 &  LMC &  2.2000 & 83.1050 & -64.7364 &      1.63$\pm$0.03 & 0.27$\pm$0.03    &  4.1 $\pm$ 0.18 &	 4.46	&  10.50 & 47.1 & 	1.44	& B & 0.39 $\pm$ 0.02 & c & -0.15 & c\\
    Hodge6 &  LMC &  2.5000 & 85.5721 & -71.5910 &      1.99$\pm$0.03 & 0.70$\pm$0.03    &  4.5 $\pm$ 0.20 &	 4.88	&  10.69 & 50.1 & 	1.61	& B & 0.31 $\pm$ 0.00 & d & -0.40 & d\\
     Kron3 &  SMC &  6.5000 &  6.1943 & -72.7936 &      0.54$\pm$0.03 & -1.42$\pm$0.03   &  5.1 $\pm$ 0.25 &	 5.62	&  14.45 & 60.6 & 	2.63	& B & 0.18 $\pm$ 0.02 & c & -1.16 & c\\
  Lindsay1 &  SMC &  7.5000 &  0.9768 & -73.4719 &      0.57$\pm$0.01 & -1.53$\pm$0.01   &  4.9 $\pm$ 0.23 &	 5.43	&  22.91 & 56.9 & 	1.8	& B & 0.18 $\pm$ 0.02 & c & -1.35 & c\\
Lindsay113 &  SMC &  5.3000 & 27.3737 & -73.7278 &      1.31$\pm$0.02 & -1.21$\pm$0.02   &  4.6 $\pm$ 0.20 &	 5.10	&  22.74 & 57.5 & 	1.27	& B & 0.18 $\pm$ 0.02 & c & -1.24 & c\\
 Lindsay38 &  SMC &  6.5000 & 12.2075 & -69.8702 &      0.57$\pm$0.05 & -0.84$\pm$0.04   &  3.9 $\pm$ 0.20 &	 4.42	&  16.80 & 66.7 & 	0.58	& B & 0.07 $\pm$ 0.00 & d & -1.50 & d\\
    NGC121 &  SMC & 10.5000 &  6.7039 & -71.5359 &      0.27$\pm$0.02 & -1.13$\pm$0.02   &  5.4 $\pm$ 0.28 &	 5.97	&  10.40 & 64.9 & 	4.49	& B & 0.18 $\pm$ 0.02 & c & -1.71 & c\\
   NGC1466 &  LMC & 12.5893 & 56.1375 & -71.6717 &      1.73$\pm$0.02 & -0.68$\pm$0.02   &  5.2 $\pm$ 0.09 &	 5.80	&   4.98 & 54.2 & 	4.92	& M & 0.31 $\pm$ 0.08 & M & -2.17 & M\\
   NGC1651 &  LMC &  2.0000 & 69.3843 & -70.5863 &      1.98$\pm$0.02 & -0.33$\pm$0.02   &  4.4 $\pm$ 0.20 &	 4.76	&  13.68 & 49.1 & 	1.21	& B & 0.35 $\pm$ 0.05 & c & -0.37 & c\\
   NGC1718 &  LMC &  1.9953 & 73.1042 & -67.0517 &      1.87$\pm$0.02 & -0.32$\pm$0.02   &  4.8 $\pm$ 0.13 &	 5.16	&   8.73 & 55.7 & 	2.05	& M & 0.26 $\pm$ 0.29 & M & -0.42 & M\\
   NGC1754 &  LMC & 13.0000 & 73.5708 & -70.4417 &      2.05$\pm$0.03 & -0.10$\pm$0.04   &  5.2 $\pm$ 0.26 &	 5.80	&   3.32 & 50.1 & 	5.88	& M & 0.30 $\pm$ 0.08 & M & -1.54 & M\\
   NGC1777 &  LMC &  1.1000 & 73.9500 & -74.2833 &      2.10$\pm$0.02 & -0.08$\pm$0.03   &  4.2 $\pm$ 0.20 &	 4.52	&  11.23 & 51.5 & 	1.26	& B & 0.39 $\pm$ 0.02 & c & -0.35 & c\\
   NGC1783 &  LMC &  1.5000 & 74.7874 & -65.9872 &      1.64$\pm$0.01 & -0.06$\pm$0.01   &  5.1 $\pm$ 0.16 &	 5.45	&  17.36 & 49.2 & 	2.44	& B & 0.30 $\pm$ 0.03 & c & -0.65 & f\\
   NGC1786 &  LMC & 13.0000 & 74.7750 & -67.7450 &      1.92$\pm$0.01 & 0.08$\pm$0.02    &  5.5 $\pm$ 0.16 &	 6.10	&   4.48 & 48.8 & 	8.43	& M & 0.39 $\pm$ 0.08 & M & -1.87 & M\\
    NGC1806 &  LMC &  1.5000 & 75.5458 & -67.9881 &      1.87$\pm$0.02 & -0.05$\pm$0.02   &  4.8 $\pm$ 0.25 &	 5.15	&  11.57 & 49.2 & 	2.16	& B & 0.25 $\pm$ 0.04 & c & -0.71 & c\\
    NGC1835 &  LMC & 13.0000 & 76.2708 & -69.4033 &      1.87$\pm$0.02 & -0.18$\pm$0.03   &  5.8 $\pm$ 0.12 &	 6.40	&   2.31 & 50.1 & 	12.22	& M & 0.33 $\pm$ 0.08 & M & -1.79 & M\\
   NGC1841 &  LMC & 12.3027 & 71.3458 & -83.9967 &      1.96$\pm$0.01 & -0.04$\pm$0.01   &  5.0 $\pm$ 0.12 &	 5.59	&  10.57 & 52.0 & 	3.68	& M & 0.66 $\pm$ 0.06 & M & -2.11 & M\\
   NGC1846 &  LMC &  1.5000 & 76.8958 & -67.4608 &      1.76$\pm$0.01 & 0.08$\pm$0.01    &  4.8 $\pm$ 0.21 &	 5.15	&  15.04 & 49.2 & 	2.02	& B & 0.41 $\pm$ 0.04 & c & -0.70 & c\\
   NGC1898 &  LMC & 13.0000 & 79.1750 & -69.6567 &      2.08$\pm$0.03 & 0.25$\pm$0.03    &  5.3 $\pm$ 0.15 &	 5.89	&  18.20 & 47.9 & 	4.79	& M & 0.26 $\pm$ 0.08 & M & -1.37 & M\\
   NGC1916 &  LMC & 12.0000 & 79.6562 & -69.4069 &      1.70$\pm$0.06 & -0.46$\pm$0.08   &  5.9 $\pm$ 0.20 &	 6.49	&   4.26 & 50.0 & 	13.83	& B & 0.42 $\pm$ 0.05 & c & -2.08 & c\\
   NGC1978 &  LMC &  2.0000 & 82.1863 & -66.2364 &      1.79$\pm$0.01 & 0.36$\pm$0.02    &  5.2 $\pm$ 0.11 &	 5.56	&  11.69 & 47.9 & 	3.3	& B & 0.76 $\pm$ 0.05 & c & -0.42 & c\\
    NGC2005 &  LMC & 13.0000 & 82.5375 & -69.7517 &      1.69$\pm$0.04 & 0.45$\pm$0.05    &  5.4 $\pm$ 0.21 &	 6.00	&   1.96 & 49.0 & 	12.05	& M & 0.37 $\pm$ 0.08 & M & -1.92 & M\\
  NGC2019 &  LMC & 13.0000 & 82.9833 & -70.1600 &      1.95$\pm$0.04 & 0.45$\pm$0.04    &  5.6 $\pm$ 0.15 &	 6.20	&   3.58 & 50.1 & 	9.58	& M & 0.43 $\pm$ 0.08 & M & -1.81 & M\\
 NGC2121 &  LMC &  3.2000 & 87.0551 & -71.4797 &      1.79$\pm$0.01 & 0.97$\pm$0.01    &  4.9 $\pm$ 0.23 &	 5.35	&  19.85 & 45.8 & 	1.88	& B & 0.53 $\pm$ 0.04 & c & -0.61 & c\\
 NGC2153 &  LMC &  1.2882 & 89.4625 & -66.4000 &      1.63$\pm$0.06 & 0.56$\pm$0.07    &  3.9 $\pm$ 0.26 &	 4.23	&   2.34 & 50.1 & 	1.09	& M & 0.09 $\pm$ 0.25 & M & -0.42 & M\\
   NGC2155 &  LMC &  3.2000 & 89.6338 & -65.4774 &      1.66$\pm$0.02 & 0.85$\pm$0.03    &  4.2 $\pm$ 0.18 &	 4.62	&  10.54 & 45.7 & 	1.02	& B & 0.43 $\pm$ 0.04 & c & -0.55 & c\\
   NGC2162 &  LMC &  1.3000 & 90.1267 & -63.7220 &      1.56$\pm$0.02 & 0.79$\pm$0.02    &  3.9 $\pm$ 0.20 &	 4.23	&  14.32 & 52.0 & 	1.12	& B & 0.39 $\pm$ 0.02 & c & -0.23 & c\\
   NGC2173 &  LMC &  1.6000 & 89.4933 & -72.9787 &      2.01$\pm$0.02 & 0.88$\pm$0.02    &  4.1 $\pm$ 0.20 &	 4.44	&   9.78 & 48.7 & 	1.07	& B & 0.39 $\pm$ 0.02 & c & -0.24 & c\\
   NGC2193 &  LMC &  2.1000 & 91.5708 & -65.0983 &      1.48$\pm$0.02 & 0.98$\pm$0.02    &  3.8 $\pm$ 0.20 &	 4.17	&  10.00 & 48.9 & 	1.44	& B & 0.39 $\pm$ 0.02 & c & -0.60 & c\\
   NGC2203 &  LMC &  1.8000 & 91.1776 & -75.4378 &      1.95$\pm$0.01 & 0.83$\pm$0.01    &  4.5 $\pm$ 0.16 &	 4.86	&  13.72 & 48.1 & 	1.38	& B & 0.39 $\pm$ 0.02 & c & -0.52 & c\\
  NGC2210 &  LMC & 12.5000 & 92.8818 & -69.1219 &      1.52$\pm$0.02 & 1.34$\pm$0.01    &  5.4 $\pm$ 0.18 & 	 5.99	&   6.00 & 47.6 & 	7.02	& B & 0.39 $\pm$ 0.02 & c & -1.97 & c\\
   NGC2213 &  LMC &  1.5849 & 92.6750 & -71.5283 &      1.81$\pm$0.02 & 0.98$\pm$0.02    &  4.5 $\pm$ 0.11 &	 4.83	&   5.50 & 51.5 & 	1.82	& M & 0.06 $\pm$ 0.12 & M & -0.01 & M\\
  NGC2231 &  LMC &  1.5800 & 95.1796 & -67.5186 &      1.49$\pm$0.04 & 1.09$\pm$0.04    &  3.9 $\pm$ 0.20 &	 4.26	&  12.65 & 50.1 & 	1.35	& B & 0.39 $\pm$ 0.02 & c & -0.67 & c\\
    NGC2257 &  LMC & 13.0000 & 97.5500 & -64.3267 &      1.40$\pm$0.01 & 0.93$\pm$0.01    &  5.3 $\pm$ 0.09 &	 5.90	&  13.24 & 52.7 & 	2.48	& M & 0.00 $\pm$ 0.00 & M & -1.63 & M\\
    NGC339 &  SMC &  6.0000 & 14.4440 & -74.4703 &      0.65$\pm$0.02 & -1.24$\pm$0.02   &  4.8 $\pm$ 0.30 &	 5.31	&  15.98 & 57.6 & 	1.77	& B & 0.18 $\pm$ 0.02 & c & -1.50 & c\\
    NGC361 &  SMC &  8.1000 & 15.5535 & -71.6045 &      0.84$\pm$0.02 & -1.31$\pm$0.02   &  5.0 $\pm$ 0.18 &	 5.54	&  12.50 & 56.8 & 	3.28	& B & 0.17 $\pm$ 0.02 & c & -1.45 & c\\
    NGC416 &  SMC &  6.0000 & 16.9965 & -72.3555 &      0.87$\pm$0.03 & -1.21$\pm$0.02   &  5.2 $\pm$ 0.26 &	 5.71	&   8.53 & 60.4 & 	4.4	& B & 0.20 $\pm$ 0.02 & c & -1.44 & c\\
    NGC419 &  SMC &  1.5000 & 17.0732 & -72.8844 &      0.86$\pm$0.02 & -1.19$\pm$0.01   &  5.0 $\pm$ 0.13 &	 5.35	&  12.57 & 58.8 & 	2.87	& B & 0.32 $\pm$ 0.02 & c & -0.60 & c\\
   Reticulum &  LMC & 12.0000 & 69.0458 & -58.8626 &      1.97$\pm$0.02 & -0.32$\pm$0.02   &  4.7 $\pm$ 0.20 &	 5.29	&  30.86 & 47.6 & 	1.1	& B & 0.14 $\pm$ 0.03 & i & -1.88 & j\\
     SL663 &  LMC &  3.2000 & 85.6212 & -65.3629 &      1.69$\pm$0.02 & 0.61$\pm$0.02    &  4.2 $\pm$ 0.19 &	 4.62	&  12.95 & 46.1 & 	0.95	& B & 0.38 $\pm$ 0.04 & c & -0.60 & c\\
      SL842 &  LMC &  2.2000 & 92.0625 & -62.9883 &      1.45$\pm$0.04 & 0.74$\pm$0.04    &  3.7 $\pm$ 0.20 &	 4.07	&  13.35 & 50.1 & 	1.09	& B & 0.39 $\pm$ 0.02 & c & -0.36 & c\\
     SL855 &  LMC &  1.3500 & 92.7208 & -65.0433 &      1.48$\pm$0.04 & 1.06$\pm$0.03    &  3.7 $\pm$ 0.20 &	 4.03	&  14.82 & 50.1 & 	0.46	& B & 0.39 $\pm$ 0.02 & c & -0.42 & c\\   
\hline
\end{tabular}
\hfill\parbox[t]{\textwidth}{Description of the columns: (1) Cluster ID, (2) Galaxy, (3) Cluster Age, (4) Right ascension, (5) Declination, (6) Proper motion in Right ascension, (7) Proper motion in Declination, (8)  Cluster log-mass, (9) Cluster initial log-mass, (10) Half-light radius,  (11) Line-of-sight distance, (12) Velocity dispersion, (13) Data source: M stands for \citet{McLaughlin2005} and B for the database of Structural Parameters of Local Group Star Clusters\footref{footnote:CatHolger}, (14) Visual extinction. (16) Metallicity, (15) and (17) Sources where the extinction and metallicity are extracted from: (a) \citet{Grebel2000}, (c) \citet{Pessev2006}, (d) \citet{Chantereu2019}, (e) \citet{Ahumada2019}, (f) \citet{Santos2004}, (g) \citet{Piatti2005}, (h) \citet{Piatti2008}, (i) \citet{Kuehn2013}, (j) \citet{Pieres2016}, (n) \citet{Schlegel1998}}
\end{center}
\label{Tab:clusters}
\end{table*}

We represent the background model likelihood by making a Gaussian Mixture Model (GMM) of the distribution of stars in a control field around the cluster. The control field is taken as an annulus with an inner radius of 3 times the cluster's half-light radius \citep{CuevasOtahola2021} and an adaptive outer radius selected such that it includes 10,000 stars, and assume the behaviour of the background in the annulus to be representative of that in the cluster area. We then estimate the background likelihood by fitting a GMM of the distribution of background stars in parallax and the two proper motion components using between 3 and 12 components and determine the optimal number of components for each cluster by minimizing the Bayesian Information Criterion (BIC). In most cases, the optimal number of GMM components found was 4.

Having the cluster and background likelihood models required by Equation~\ref{eq:L_mixture_model}, and having marginalised over the cluster's parameters $\theta_{cl}$, \neww{the inference problem at this stage reduces to inferring the weight parameter $f$ (for each cluster)}. Since the number of RRL stars expected to belong to each cluster is small by the very nature of the problem, we infer the mixture weight $f$ using all stars (not just RRL stars) in Gaia DR3 \citep{GaiaCol_DR3}, from the posterior probability density function

\begin{equation}
    p(f|\{D_n\}) \propto p(f) \prod _n^N p(D_n| f)
\end{equation}

\begin{figure*}
\includegraphics[width=\columnwidth]{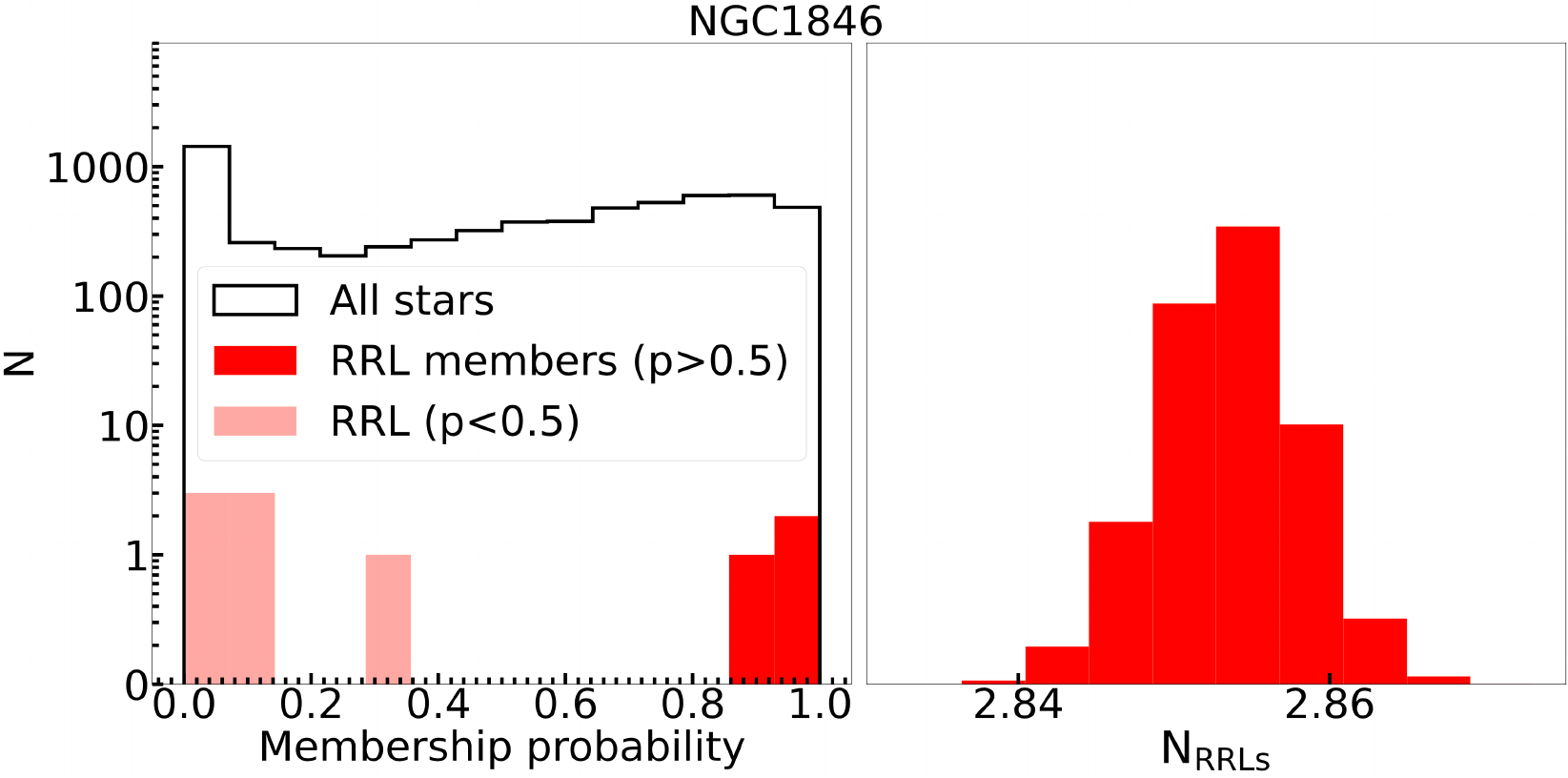}
\includegraphics[width=\columnwidth]{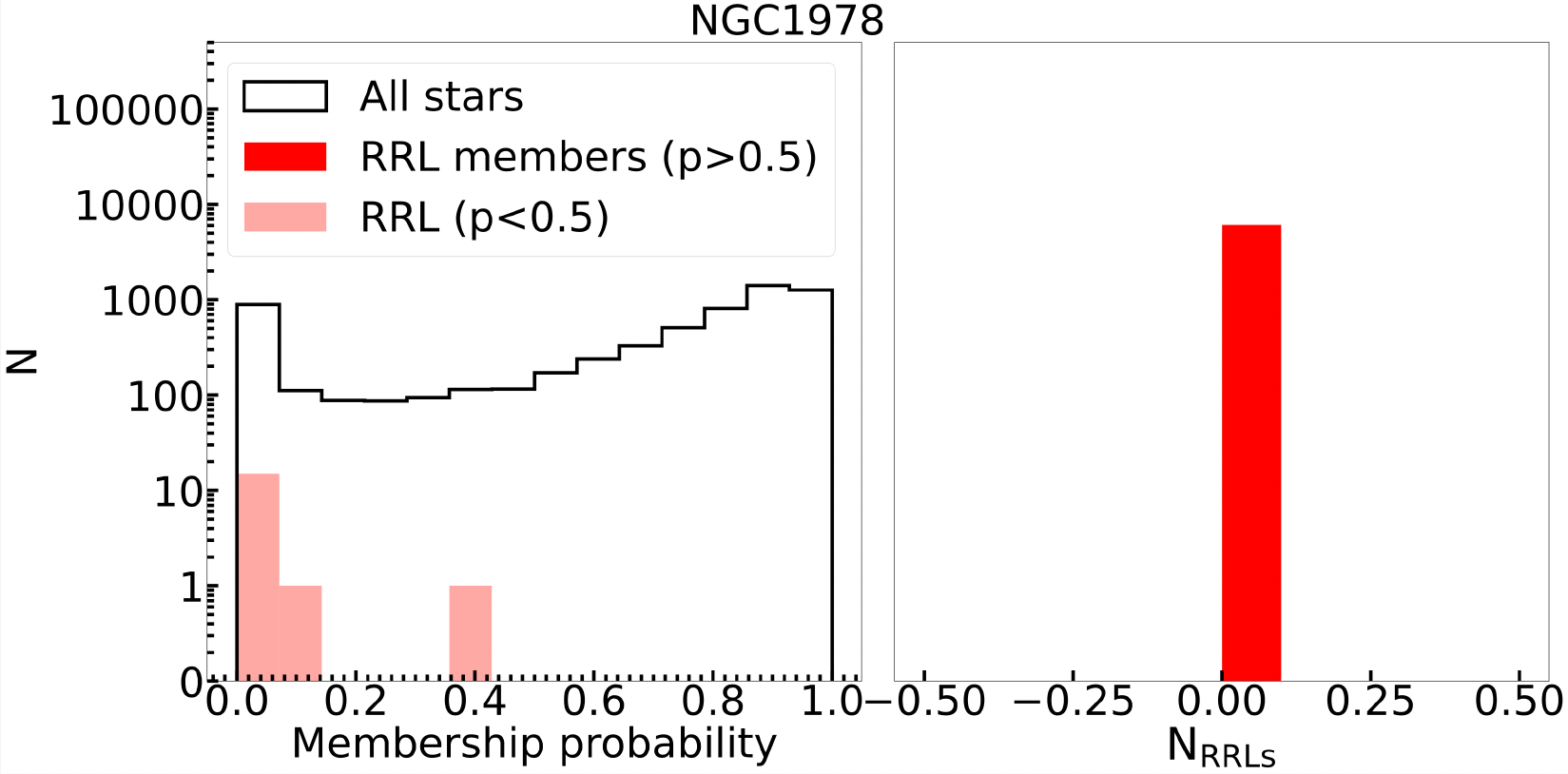}
\caption{Posterior distribution of the inferred number of RR Lyrae in the clusters NGC~1846 (\emph{left}) and NGC~1978 (\emph{right}). The solid (black) line represents the histogram for all stars (generic and RRL), the dark red histograms correspond to all RRL  within 3$r_h$ and probable cluster RRL members ($p>0.5$ and $<3R_h$), whereas the light red ones correspond to RRL with low membership probability  RRLs ($p<0.5$)}
\label{fig:post_dist_nrrls_NGC_1846}
\end{figure*}

\begin{figure*}
  \centering
\includegraphics[width=\columnwidth]{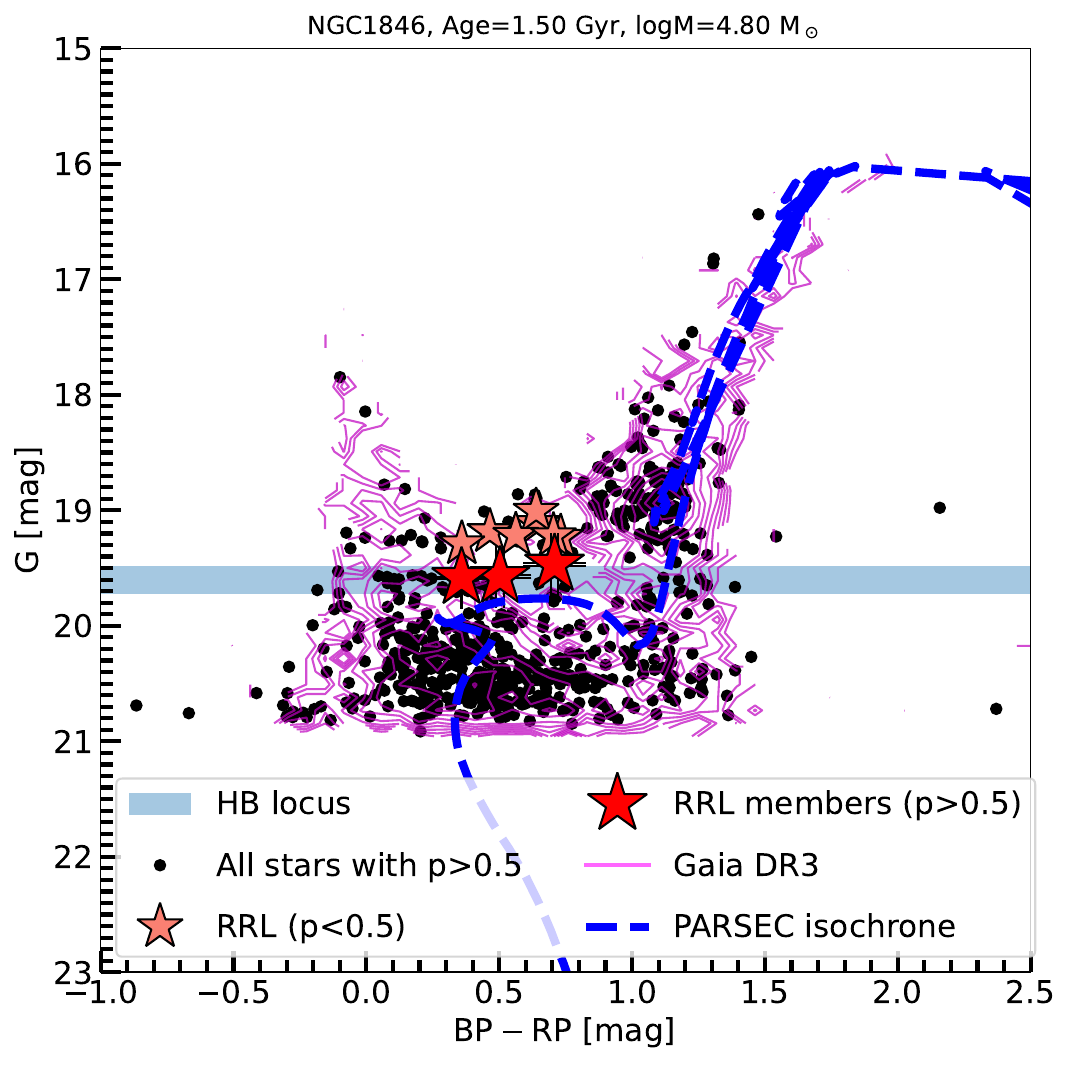}
\includegraphics[width=\columnwidth]{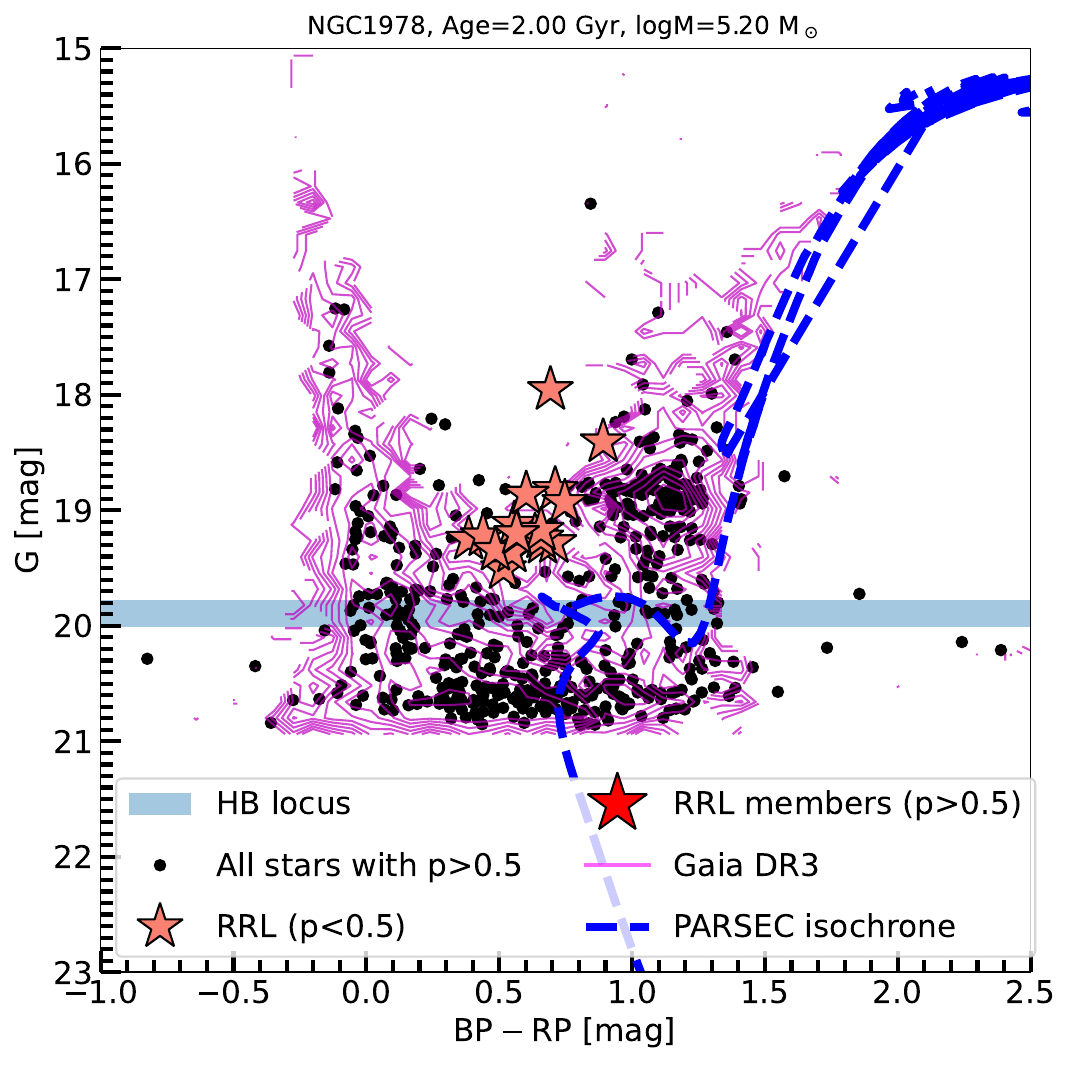}
\caption{$G$ vs $BP-RP$ color magnitude diagrams for stars in the fields around NGC~1846 (left) and NGC~1978 (right). The background contours show the number counts of all Gaia~DR3 stars in each field. Black dots correspond to generic stars with high probabilities ($p>0.5$) that would be classified as members. RRL members and non-members are shown with five-point stars in dark and light red respectively. The apparent magnitude range corresponding to each cluster's horizontal branch is shown by the blue stripe (width corresponds to $\pm$0.15~mag = $\pm$7\% uncertainty in distance).} 
\label{fig:CMD_NGC_1846}
\end{figure*}

assuming for $f$ a uniform prior $p(f)=U(0,1)$. To generate posterior samples for $f$, we used the Markov Chain Monte Carlo sampler \verb|emcee| \citep{ForemanMackey2013_emcee_PASP,ForemanMackey2013_emcee_ASCL}, running chains with 32 walkers, first for 500 burn-in steps, and restarted for another 4000 steps.  The burn-in and final number of steps were estimated based on initial runs which yielded a maximum auto-correlation time, and rerun such that final runs were respectively $>10$ (burn-in) and $>100$ times the autocorrelation time.

Following \citet{ForemanMackey2014_blog}, the membership probability of the $n$-th star to be a member of the cluster is given by:

\begin{equation}
    p_n = \frac{f\mathcal{L}_{cl}}{f\mathcal{L}_{cl} + (1-f)\mathcal{L}_{bg}} \label{eq:membership_prob}
\end{equation}

and for each star the mean membership probability is computed along the chain, thinned by the auto-correlation time of each chain (usually $\sim 30$). Finally, we take the RRL stars with mean membership probabilities $p>0.5$ and within an angular distance $\Delta<3R_h$ as probable cluster members. As some clusters in Tab. \ref{Tab:clusters} have low $R_h$ values (less than 5 pc), the latter limit does not enclose the majority of stars, which in the case of young clusters is due to clusters not tidally limited by dynamical evolution. For some older and more extended clusters, such as NGC 1466, we carefully examined the values in the literature, and to illustrate this, we considered as the upper limit in angular distance its tidal radius $R_t=$~40.5~pc \citep{Lanzoni_2019}, which is considerably larger that $3R_h=14.94$~pc. For this reason, we consider $R_t$ as an upper limit in angular distance for clusters with low $R_h$ values, which ensures that we consider stars still bound to the cluster. 

We also compute  the expected number of RRL members $\langle N^{RRLs}_i \rangle$ as the sum of the membership probabilities of all RRL stars in a given cluster's field. Since this is well-defined for each (independent) step of the chain we can also obtain $P(N^{RRLs}_i|{D_n}_i)$ a posterior distribution for the inferred number of RRL stars in the $i$-th cluster. 
Figure~\ref{fig:post_dist_nrrls_NGC_1846} illustrates the distributions of membership probabilities (left) and of the number of members $N_{RRLs}$ (right) for two representative cases: NGC~1846 (top), a cluster with 3 probable cluster members, and NGC~1978 (bottom), a case with none. The distribution of membership probabilities is bimodal with the majority of stars likely belonging to the background, as expected.

Figure~\ref{fig:CMD_NGC_1846} shows the distribution of RRL and non-RRL stars and members in the $G$ vs $BP-RP$ color-magnitude diagram (CMD), again for NGC 1846 and NGC 1978 (left and right panels, respectively). The HB locus at each cluster's distance is shown with a light blue stripe. We use star symbols to indicate those stars in Gaia DR3 contained in the master catalogue and classified as RRLs, with dark and light and red symbols corresponding respectively to RRLs classified as likely cluster members ($p>0.5$ and $\Delta<3R_h$) and non-members, respectively. We notice that in the left panel of Fig. \ref{fig:CMD_NGC_1846} the three stars with the highest membership probability lie within the HB locus, supporting that those RRLs are likely members of the NGC 1846 cluster. In the right panel, on the other hand,  all stars in the CMD diagram lie outside the HB and have low membership probabilities, as expected. The CMDs in both panels show the member samples of generic stars are indeed contaminated, as evidence clearly by the conspicuous Red Clump stars identified as members whose apparent magnitude is inconsistent with the cluster's distance: if they were true cluster members, these stars should lie along the HB. Because the membership model is using parallax information, significantly uncertain for generic stars at these distances, it incorrectly finds them as likely cluster members. For the RRL the situation is much better thanks to their much more precise photometric distances, and hence photometric parallaxes, which make the parallax likelihood term more informative and able to better discriminate background RRLs at similar distances as the Red Clump stars that would otherwise have been mistaken for members (light red stars). From this we conclude that the inferred membership probabilities of RRLs are much more robust that those for generic stars and our final member sample is less prone to contamination, although probably not completely free from it.

\begin{table*}
\setlength{\tabcolsep}{5 pt}
\small
\caption{Cluster member candidate RRL stars}
\label{Tab:cands}
\begin{tabular}{lllccccccclcc}
\hline
Cluster & Galaxy & Gaia Source ID & RA & DEC &  G & $\rm AmpV$ & Period & Type &  p & $\Delta/R_{lim}$ & [Fe/H]$_{Li23}$ \\
 & & &  (deg) & (deg) &  (mag) & (mag)  & (d) & & & \\
(1) & (2) & (3) & (4) & (5) & (6) & (7) & (8) & (9) & (10) & (11) & (12)\\
\hline
NGC1466 & LMC & 4641999382207275008 & 56.2041432 & -71.6803283 & 19.33 & 0.43 & 0.348587 & RRC & 0.968 & 0.53 & -- \\
NGC1466 & LMC & 4641999386501750016 & 56.1962819 & -71.6778392 & 19.29 & 0.48 & 0.616186 & RRAB & 0.999 &  0.45 & -- \\
NGC1466 & LMC & 4642001929123686528 & 56.1277290 & -71.6898080 & 19.33 & 1.03 & 0.590836 & RRAB & 0.998 &   0.43  & -1.85\\
$\cdots$ & $\cdots$ & $\cdots$ & $\cdots$ & $\cdots$ & $\cdots$ & $\cdots$ & $\cdots$ & $\cdots$ & $\cdots$ & $\cdots$ & $\cdots$ \\
\hline
\end{tabular} 
\hfill\parbox[t]{\textwidth}{Description of the columns: (1) Cluster the candidates are associated to, (2) Galaxy, (3) Gaia DR3 source ID of each candidate, (4) Right ascension, (5) Declination, (6) Mean Gaia G band magnitude, (7) OGLE V-band amplitude, (8) Candidates orbital period, (9) Candidates type, (10) Membership probability computed using the inference model we described in this work, (11) Ratio of the star's angular distance to the upper limit of the cluster ($R_t$ or $3R_h$), (12) \newww{Photometric metallicity from \citet{Li2023}}. This table is shown in its entirety in the electronic edition. A portion is shown here for guidance.}
\end{table*}

Table~\ref{Tab:cands} summarises the list of identified RRL cluster  members candidates ($p>0.5$, $\Delta<3R_h$) including their membership probabilities and general astrometric and light curve properties. Figure~\ref{fig:ap_pm_and_pos} in
Appendix~\ref{Ap:pm_and_pos} shows the distribution in the sky and proper motion planes for the RRL member (and non-member) candidates around intermediate-age clusters. 
Figure~\ref{fig:lcs} in Appendix~\ref{Ap:lcs} shows the G-band light curves for the \NRRintProb\ RRL identified as member candidates of intermediate-age clusters ( ages from 1 to 8 Gyr), as well as the Period-Amplitude diagram for these and all RRL stars in Table~\ref{Tab:cands}.

\begin{figure}
    \centering
    \includegraphics[width=\columnwidth]{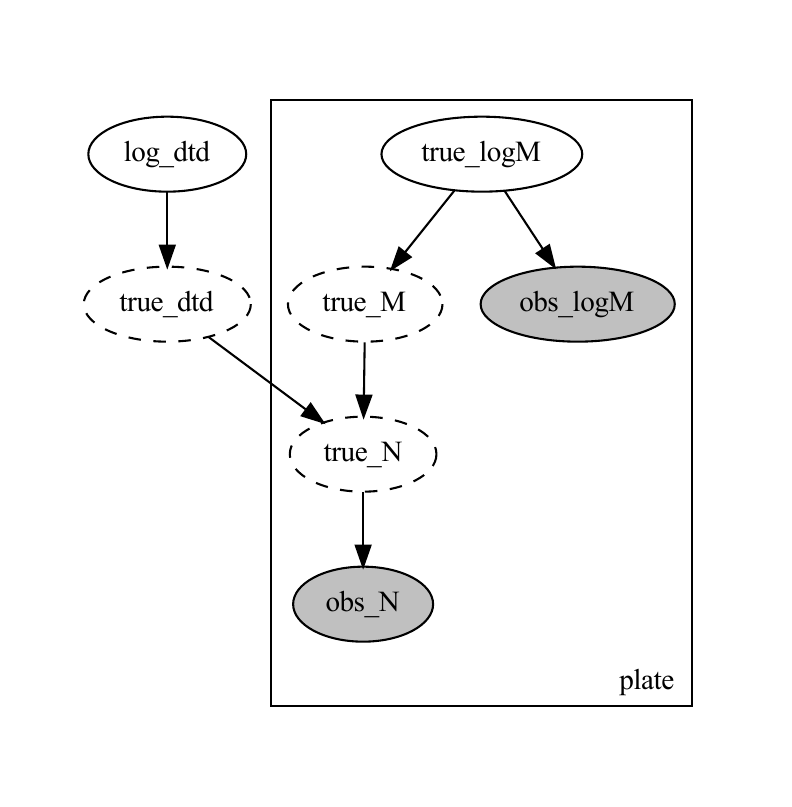}
    \caption{Graphical representation of the relations between parameters in the probabilistic hierarchical inference model used for the DTD. Observed data are represented with grey ovals. The rectangle encloses the part of the model specific to each cluster, while population-wide parameters (in this case the DTD) are outside of it. }
    \label{fig:model_graph}
\end{figure}

\section{Inference of the delay time distribution}\label{Sec:dtd_inference_model}

The DTD stands for the number of a specific object type, in this case RRL stars, per unit stellar mass formed during a burst of star formation at a given look-back time (or age). Preserving the notation from S21, the number $\lambda$ of RRLs present in the population at a given time can be expressed as 
 
\begin{equation}
    \lambda=\sum_{j=1}^J \lambda_j =\sum_{j=1}^J {M_{j}} DTD_j
\label{Eq:lambda}
\end{equation}

with $M_j$ the stellar mass formed at an age $j$; $\DTD_j$, the DTD at that age; and the sum going over the ages available in the star formation history (or stellar age distribution). 

Assuming each cluster to be a simple stellar population of a given age, we infer the posterior probability for the $\logDTD_j$ at a given age $j$ based on the number of RRLs per cluster as:

\begin{multline}
    P(\logDTD_j| \{N^{obs},M^{obs}\}) = \\ 
    \prod_{i=1}^{ N_\mathrm{clusters} } P_i(N^{obs},{M^{obs}}|\logDTD_j)P(\logDTD_j)
\label{Eq:p_ensemble}
\end{multline}

where $P(\logDTD_j)$ is the prior for the DTD and $P_i(\logDTD_j|N^{obs},M^{obs})$ is likelihood of the observed number of RRL stars $N^{obs}$ and \emph{initial} cluster mass $M$ (from Tables~\ref{Tab:NRR_per_cluster} and \ref{Tab:clusters}, respectively), given the $\DTD$ at age $j$ for the $i$-th cluster. We have chosen to subscript the probability $P$ to simplify the notation to indicate all (observed) quantities refer to the $i$-th cluster. 

\begin{table}
\caption{Number of RRL members per cluster}
\centering
\label{Tab:NRR_per_cluster}
\begin{tabular}{|l|l|c|c|c|}
\hline
\multicolumn{1}{|l|}{IDs} & \multicolumn{1}{|l|}{Age} & \multicolumn{1}{c|}{$\rm N_{RRLs}$} & \multicolumn{1}{c|}{$\rm \langle N_{RRLs}\rangle$} & \multicolumn{1}{c|}{$N_\mathrm{cont}$}  \\
& (Gyr) & \multicolumn{1}{c|}{$\rm (p>0.5)$} & (all) &  \\
(1) & (2) & (3) & (4) & (5)\\
\hline
NGC1777         &  1.1 &  0 &    0 & 0.0 \\
NGC2153         &  1.3 &  2 &    2 & 1.7 \\
NGC2162         &  1.3 &  0 & 0.00043 & 0.0 \\
SL855           &  1.4 &  0 & 6.6$\times 10^{-5}$ & 0.0 \\
NGC1846         &  1.5 &  3 &  2.9 & 1.3 \\
NGC1806         &  1.5 &  2 &  1.7 & 1.7 \\
NGC419          &  1.5 &  1 & 0.93 & 0.7 \\
NGC1783         &  1.5 &  0 & 4.4$\times 10^{-5}$ & 1.1 \\
NGC2231         &  1.6 &  0 & 0.0038 & 0.0 \\
NGC2213         &  1.6 &  3 &    3 & 1.6 \\
NGC2173         &  1.6 &  0 &    0 & 0.3 \\
NGC2203         &  1.8 &  0 &    0 & 0.1 \\
Hodge14         &  1.8 &  3 &    3 & 1.8 \\
NGC1718         &  2.0 &  1 &    1 & 1.3 \\
NGC1651         &  2.0 &  0 &    0 & 0.8 \\
NGC1978         &  2.0 &  0 & 0.37 & 0.1 \\
NGC2193         &  2.1 &  0 &    0 & 0.0 \\
SL842           &  2.2 &  0 &    0 & 0.1 \\
Hodge4          &  2.2 &  0 &    0 & 0.2 \\
Hodge6          &  2.5 &  0 &    0 & 1.1 \\
SL663           &  3.2 &  0 &    0 & 0.2 \\
NGC2121         &  3.2 &  3 &  3.5 & 4.0 \\
NGC2155         &  3.2 &  0 & 0.09 & 0.3 \\
Lindsay113      &  5.3 &  0 &    0 & 0.2 \\
NGC339          &  6.0 &  3 &    3 & 1.3 \\
NGC416          &  6.0 &  0 &    0 & 0.2 \\
Lindsay38       &  6.5 &  0 &    0 & 0.1 \\
Kron3           &  6.5 &  0 &    0 & 0.6 \\
Lindsay1        &  7.5 &  0 &    0 & 0.3 \\
NGC361          &  8.1 &  1 &    1 & 1.0 \\
NGC121          & 10.5 &  5 &  4.8 & 0.1 \\
Reticulum       & 12.0 & 23 &   24 & 0.2 \\
NGC1841         & 12.3 &  5 &  5.8 & 0.0 \\
NGC2210         & 12.5 & 17 &   17 & 0.2 \\
NGC1466         & 12.6 & 33 &   34 & 0.0 \\
NGC1754         & 13.0 & 12 &   12 & 4.5 \\
NGC1786         & 13.0 & 25 &   25 & 3.1 \\
NGC2257         & 13.0 & 31 &   31 & 0.2 \\
Hodge11         & 13.0 &  1 & 0.99 & 1.6 \\
\hline
\end{tabular}
\hfill\parbox[t]{\columnwidth}{Description of the columns: (1) Cluster (2) Age, (3) Number of RRL members with $p>0.5$ and $r<3R_h$ (4) Expectation value of the number of RRL members ($\sum p$ for RRLs with within $r<3R_h$) (5) Median number of contaminants estimated in 8 control fields.}
\end{table}

The likelihood for the DTD of the $i$-th cluster is in turn given by

\begin{multline}
P_i(N^{obs},M^{obs}|DTD) \\
    =\int P(N^{obs}|M,DTD)P(M^{obs}|M)P(M) dM
\label{Eq:p_each}
\end{multline}

where $P(N^{obs}|M,DTD)$ is the term that models the likelihood of having observed $N^{obs}$ number of RRLs given the expected the $\logDTD_j$ and the \emph{true} cluster mass $M$; multiplying this by the last two terms $P(M^{obs}|M)P(M)$ and marginalising over the mass takes into account the uncertainty in the true mass of the cluster. This is a simple hierarchical model in which the inference of a population-wide parameter, in this case the DTD, affecting all clusters equally, depends on the modelling of one or more parameters for each cluster, in this case the mass. Figure~\ref{fig:model_graph} shows a graphical representation of the interdependence of the different model parameters.

The first term in the likelihood in Eq.~\ref{Eq:p_each} is modelled as a Poisson process by means of the Gamma distribution, a generalisation of the Poisson distribution for real valued $N^{obs}$, expressed as

\begin{equation}
P(N^{obs}|M_j,DTD_j) = \frac{\beta^\alpha}{\Gamma(\alpha)} (N^{obs})^{\alpha-1} e^{\beta N^{obs}} 
\label{Eq:poisson_term}
\end{equation}

where $\alpha=\lambda_j+1=(M_j*DTD_j)+1$, i.e. the true number of RRLs given the true cluster mass $M_j$ and $DTD_j$ at age $j$, and $\beta=1$, which reduces the Gamma distribution to the Poisson distribution for integer $\alpha$. 

The second term in Eq.~\ref{Eq:p_each}, i.e. the likelihood of the observed cluster mass, $P(M^{obs}|M)$  is modelled as a Gaussian distribution with mean equal to the observed (initial) cluster mass and a standard deviation $\sigma^{obs}_{M}$ corresponding to its error as

\begin{equation}
P_i(M^{obs}|M,\sigma_{M})=\mathcal{N}(M^{obs},\sigma^{obs}_{M})
\label{Eq:mass_term}
\end{equation}

and $P(M)$, the prior for the cluster's initial mass, is assumed to be uniform in $\log{(M/M_\odot)}$ in the range $[3,7]$. Finally, we assume a uniform prior for $\logDTD(\mathrm{RRL}/M_\odot)$ in the range $[-7,-3]$. In the model we take $N^{obs}$ to be the obtained expectation value $\langle N_{RRLs} \rangle$ of RRL cluster members reported in Table~\ref{Tab:NRR_per_cluster} for each cluster.

The hierarchical inference model for the DTD was implemented using an MCMC sampler in \verb|numpyro| with 42 walkers, 1000 initial burn-in steps and 5000 steps for the final chain, with a target acceptance probability of 0.95.

\begin{figure}
  \centering
\includegraphics[width=\columnwidth]{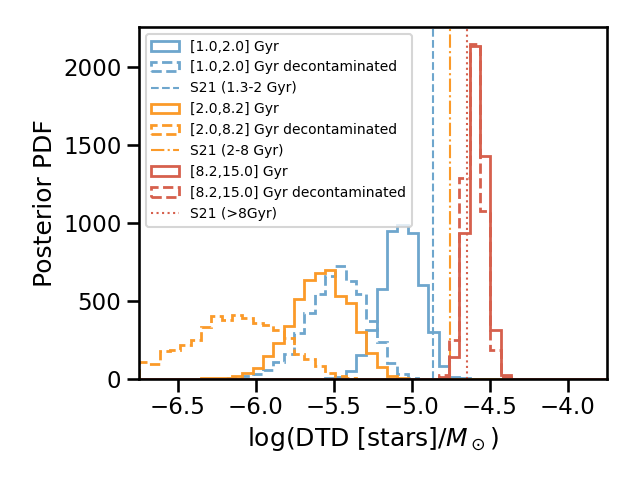}
\includegraphics[width=\columnwidth]{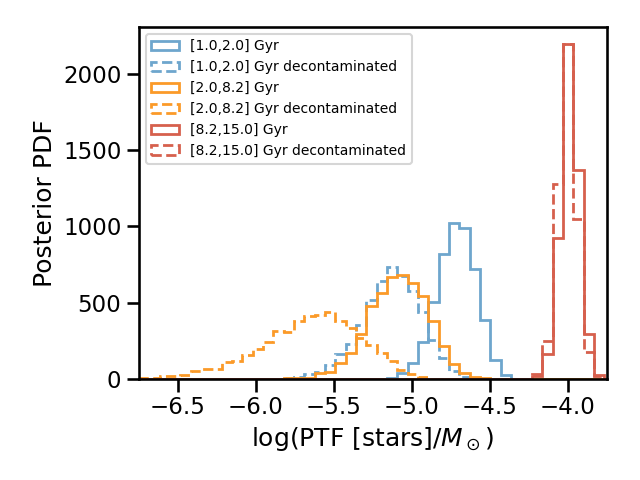}
\caption{Posterior PDF for the Delay Time Distribution (DTD \emph{top}) and for the Present Time Frequency (PTF, \emph{bottom}) of RRL stars, for clusters in three age ranges: 1 to 2 Gyr, 2 to 8 Gyr and 8 to 15 Gyr. The solid and dashed lines correspond to the posterior PDFs before (solid) and after decontamination (dashed). Mean values obtained by S21 for the DTD in the corresponding age ranges are shown with dashed, dashed dotted and dotted lines with the respective colours.}
\label{fig:DTD_PTF}
\end{figure}

\subsection{Assessing Contamination}

\neww{The inference described in the previous sections already takes the background into account, therefore, the probability of membership obtained should already reflect the degree of confidence in each star to be a cluster member. Nevertheless, because proper motion uncertainties at these distances remain relatively large and there are areas in both the LMC and SMC where the background presents strong spatial variations, we aim to provide an independent assessment of the probable contamination level in our member samples by applying our inference model to off-cluster control fields, where the number of members obtained should (ideally) be strictly zero. The mean number of RRL members identified in control fields can then be used as a measure of a more realistic level of the contamination expected in our sample and used to further correct our estimates of the DTD and PTF (Present Time Frequency). This information provides an added criterion to prioritise following up RRL members in clusters where the contamination is expected to be lowest compared to the number of identified members.   \\ 
We use the same methodology as described in the previous section to identify cluster members, now applied to eight control fields with centers at a distance of $23r_h$ north/south and/or east/west of each cluster's center. This ensures there are no cluster stars either in the area where members will be selected ($<3r_h$) nor in the background annulus (radii between $3$ and $20r_h$, the typical inner and out radii of the background annuli). All cluster parameters used for the inference are kept unchanged. RRLs identified in these control field as 'members', i.e. membership probability $p>0.5$ and within $3r_h$ of the center, are clearly false positives. The number of contaminant RRLs identified in this way, $N_{cont}$, is tallied in the last column of Table~\ref{Tab:NRR_per_cluster} and was subtracted from $\langle N_\mathrm{RRLs} \rangle$ to infer the de-contaminated DTD. 
}

\section{Results and Discussion}\label{Sec:discussion}

\subsection{The DTD}

The top panel of Figure~\ref{fig:DTD_PTF} shows the posterior probability density function (PDF) for the $\logDTD$ inferred by grouping the clusters into three age ranges: 1 to 2 Gyr, 2 to 8 Gyr and 8 to 15 Gyr with 16, 14 and 9 clusters in each range, respectively. 

The current observed frequency of RRLs per \emph{current} cluster mass is also a useful quantity and can be obtained by applying the same inference model but using the present-day cluster mass, instead of the initial mass used to infer the DTD. We dub this PTF (Present Time Frequency) and is shown in the bottom panel of Figure~\ref{fig:DTD_PTF}. Table~\ref{tab:DTD_PTF} summarises the results obtained for the DTD and PTF in each age range (only results for the ensemble of clusters are reported [solid lines in the figure]), corresponding to the median with reported uncertainties corresponding to the 16th and 84th percentiles of the posterior samples.

\begin{table}
\caption{Median DTD and PTF inferred from Magellanic cluster member RRL stars. Uncertainties  correspond to 16th and 84th percentiles.}
    \label{tab:DTD_PTF}
    \centering
    \begin{tabular}{lccc}
    \hline
                & 1-2 Gyr & 2-8 Gyr & 8-15Gyr \\ 
    \hline
DTD (RRL/$10^5 M_\odot$) &
$0.87^{+0.28}_{-0.22}$ & $0.27^{+0.14}_{-0.094}$ & $2.6^{+0.38}_{-0.32}$ \\
decontaminated &
$0.34^{+0.17}_{-0.12}$ & $0.071^{+0.073}_{-0.041}$ & $2.5^{+0.36}_{-0.31}$ \\
PTF (RRL/$10^5 M_\odot$) &
$1.9^{+0.63}_{-0.49}$ & $0.83^{+0.43}_{-0.29}$ & $10^{+1.5}_{-1.3}$ \\
decontaminated &
$0.75^{+0.38}_{-0.27}$ & $0.24^{+0.22}_{-0.13}$ & $9.9^{+1.4}_{-1.3}$ \\
    
    \hline
    \end{tabular}
    
\end{table}

The results of Figure~\ref{fig:DTD_PTF} (top) show the largest value and most precise determination for the (decontaminated) DTD in the oldest age range ($>$8 Gyr) at $2.5\times10^{-5}$ RRL$/M_\odot$. This is expected since, as discussed in the introduction (Sec.~\ref{Sec:intro}), RRL stars are most abundant in stellar populations with ages $\geq 10$~Gyr and LMC/SMC clusters have relatively low metallicities ($\FeH<-0.5$), with most being precisely in the range in which the production of RRL stars is most efficient \citep[][see their Fig.~16]{CruzReyes2024}. \neww{This value is also the least affected by contamination.} 
The (decontaminated) DTD for the young (1-2 Gyr) and intermediate age (2-8 Gyr) ranges are, respectively, about 1. and 0.5~dex lower compared to the oldest bin, their estimates are more uncertain as well, which is to be expected given their lower masses and the low number of RRLs found in clusters this age. 
It is interesting to note the young bin is found to have a larger DTD than the intermediate age. From a qualitative theoretical standpoint if binary evolution is indeed responsible for the production of these stars, an increase of the DTD as age decreases may be related to the dependence of binary fraction with the mass of the primary, observed to increase with mass \citep[see, e.g., review in][]{DucheneKraus2013}.   A quantitative analysis would be warranted to confirm whether this dependence may have an appreciable effect, but also to explore other possible explanations.
So far there are no previous results from simulations that could guide our expectations. The works of \citet{Bobrick2022} and \citet{Karczmarek2017}, who have explored the production of RRL via binary evolution using stellar synthesis models, have not reported either a DTD (or a PTF) for their simulated binary evolution RRL pulsators.

From the observational standpoint, the work of S21 discussed in Sec.~\ref{Sec:intro} offers the only available estimate of the DTD for RRLs, until now. The vertical dashed, dashed dotted and dotted lines in Fig.~\ref{fig:DTD_PTF} (top) show the values reported by S21 for RRL in the age range 1.3-2 Gyr and the means for the DTD in the 2-8 Gyr and >8 Gyr ranges. Typical uncertainties (not shown) are around 0.2~dex in all cases. The agreement in the oldest bin is remarkable and a non-trivial result since our analysis is complementary to theirs.  To infer the DTD, S21 used the LMC field population, after excising known stellar clusters. Here we have done the opposite. Our results show the DTD inferred in this work for the oldest stellar clusters ($>8$~Gyr), is similar to that for the field, inferred from S21 in the same age range (for more details see their Table~1).  For the young and intermediate-age bins, however, our work yielded systematically lower values of the DTD compared to S21, \neww{even when compared against our `raw' estimates, i.e. before decontamination}.  
For the young and intermediate-age bin, our results completely disagree with our decontaminated DTD being lower by almost 1~dex.  The disagreement in our results for these ages, where the signal is both seen and expected to be lower, could be a mix of two factors: the known incompleteness of RRL catalogues in the centre of clusters (see below) and MW field contamination having inflated the DTD in the S21 inference, to a lesser extent. Although MW contaminants were filtered out in S21, this was based on information provided in the OGLE-IV catalogue which predated Gaia DR2 and was based on distance information alone. The use of Gaia DR3 proper motions together with the probabilistic modelling of the background in our membership model \neww{together with our tests in control fields} must have a better performance filtering out contaminants, particularly for MW interlopers that are easier to filter out than the Magellanic Cloud's field population. \neww{However, this contamination would be expected to have affected S21 in all age bins, contrary to the agreement found in the oldest age bin; and throughout the entire LMC, also inconsistent with S21's findings that the LMC disc regions are where most of the intermediate-age signal of the DTD is required.}

For future work it will be relevant to robustly confirm or discard any potential differences between the cluster and field DTDs since this will likely be connected to the effect of environment density on the formation mechanism of these stars. For instance, compatible estimates of the DTD for cluster and field stars could suggest the mechanism driving envelope mass loss is not significantly affected by dynamical interactions in dense stellar systems. This could imply that the mass loss occurred via canonical winds during the giant phase, a process likely unaffected by dynamical interactions. Conversely, if the estimates are notably different, this might indicate that these binaries are preferentially formed or disrupted in dense environments. Our results compared to S21's would seem to favour the latter, with the caveats mentioned previously about possible contamination in S21's sample and the relatively low numbers in which ours are based.

Two other, opposing, effects may influence our results: contamination from the LMC/SMC field populations and incompleteness in the RRL catalogues in the central parts of clusters. Although Gaia DR3 proper motion precision is useful to differentiate MW background and foreground contaminants, telling apart each cluster from its host galaxy is much more demanding \neww{as our control field results show}. In our inference, this is mostly being done by the distance/parallax term in the likelihood thanks to the high precision of RRL photometric distances, as discussed in Sec.~\ref{Sec:model} and Figure~\ref{fig:CMD_NGC_1846}. The fact we are interested in stellar clusters and these are highly spatially concentrated systems, make the distance test performed by the likelihood a much more stringent test to pass by the more extended populations of the LMC/SMC field, filtering out \neww{contaminants as best as current uncertainties allow}, although statistically some remain. As proper motion precision increases with the next Gaia Data Release(s) membership probabilities will become more robust. 

Radial velocities would help confirm candidates and weed out contaminants, but sufficient precision of $\lesssim$ 1~km/s will be required for the cluster membership confirmation to be reliable. This precision threshold is driven by the typical velocity dispersion of clusters, which extends up to a few km/s. Such precision is achievable within reasonable exposure times using medium resolution spectrographs on 8m-class telescopes. However, membership determination can be challenging in cases where the radial velocity distribution of cluster members significantly overlaps with the broader velocity distribution of field stars. \newww{Metallicity information will also be a helpful discriminant. In the mean time, since we have not used photometric metallicities in our probability membership inference, the information  is provided in Table~\ref{Tab:cands} where available and, together with the expected number of contaminants, can help prioritize the candidates for spectroscopic follow-up.}

The other factor is completeness of the Gaia and OGLE RRL catalogues, known to be lower in the central parts of the clusters due to crowding \neww{\citep[see][for each, respectively]{CruzReyes2024,Giusti2024,Kerber2018}}. For Gaia, in particular, this problem is expected to decrease significantly even as soon as DR4, as already demonstrated by the new processing strategy used in the $\omega$Cen globular cluster in the Gaia Focused Product Release of 2023 \citep{GaiaCol2023_FPR_Weingrill}. While field contamination tends to inflate the inferred DTD, reduced completeness acts in the opposite direction. 
\neww{We have quantitatively accounted for the effect of contamination (from our control field tests) but not of incompleteness, much harder to assess. Our decontaminated results, therefore, should be taken as lower limits of the expected cluster DTD and PTF at each age.}

The bottom panel of Figure~\ref{fig:DTD_PTF} shows our inference of the PTF or the current occurrence rate of RRLs per unit mass. This is a useful quantity since it arises from present-day observables and can be compared to the occurrence rates of RRL stars in MW clusters and other systems, serving as a validation test. As expected, the difference between the DTD and PTF is more noticeable in the oldest bin, where the difference between the initial and present cluster mass is largest. For the young and intermediate-age bins there are no MW cluster counterparts to compare against, except to note that all clusters in the MW with ages less than 8~Gyr have current masses under $10^4 M_\odot$ and are not known to host any RRL stars, in agreement with upper limit predictions of $0.2$ and $0.08$ RRLs that result from our inferred PTF values for clusters in the young and intermediate-age ranges, respectively. For the oldest range, however, where there are numerous observations of RRLs associated to known Galactic globular clusters, the obtained rate of $1.0\pm0.1$ RRL$/10^{4} M_\odot$ is precisely the typical mean frequency observed in globular clusters (see Sec. 4.3 in S21). For the MW globular cluster's results the effects of contamination are shown to be negligible.

\subsection{Cluster member candidates}

The membership probability model detailed in Sec.~\ref{Sec:model} led us to identify 259 RRL stars as probable members ($p>0.5$) of 40 clusters in the LMC and SMC. Out of these, \NRRintProb\ RRL stars were identified as probable members of 10 intermediate age (1 to 8 Gyr) clusters, namely:  NGC~339, NGC~361, NGC~419 from the SMC and Hodge 14, NGC~1718, NGC~1806, NGC~1846, NGC~2121, NGC~2153, NGC~2213 from the LMC. 

Several variable star searches have been conducted in LMC and SMC clusters so far. All searches that have found RRL stars were focused around old clusters, e.g. \citet{Nemec2009} and the works by \citet{Kuehn2011,Kuehn2012,Kuehn2013} who identified 49 and 32, 53, and 31 RRL stars around the LMC clusters NGC~1466, Reticulum, NGC~1786 and NGC~2257, respectively. We found fewer RRLs stars associated to these clusters than these previous works: 33, 23, 25 and 31 RRL respectively. This, however, is reasonable due to the reduced completeness expected from Gaia \neww{at the very central regions of globular clusters \citep[see e.g.][]{CruzReyes2024}, although note that all of these works predate Gaia DR2 and, not having any kinematic membership criteria, are likely to include back/foreground MC  contaminant RRLs.}

Searches focused on younger clusters have aimed at identifying Cepheids and usually have not reached limiting magnitudes deep enough for RRL stars to be found. This is the case of \citet{Sebo1994,Welch1993} and \citet{Gieren2000} who searched for variable stars in NGC~330, NGC~2164 and NGC~1866, respectively. We have found no RRLs in NGC~1866; NGC~2164 is located in the bar region and was discarded and NGC~330 (30~Myr) being below the age threshold of 1~Gyr, was not considered in our analysis.

To the best of our knowledge, the only search for variable stars around  young/intermediate-age clusters to have found RRL stars is that of \citet{Salinas2018}. The authors used Gemini South photometry to search for variables in the LMC cluster NGC~1846 and found 77 variables, out of which 5 are RRLs: a new one plus 4 previously reported by OGLE. Given the age of the cluster (2~Gyr), these RRL were assumed by the authors to be LMC field stars (see their Sec.~3.1). Our analysis yielded 3 of those RRL stars as probable members of NGC~1846, making it one of our most promising clusters. 

Finally, the early work of \citet{Walker1989} deserves special mention. These authors searched specifically for RRL stars in the SMC clusters NGC 121 (10.5~Gyr), where four RRL were found, and the intermediate-age clusters Lindsay 1, Kron 3, Kron 7, and Kron 44, where none were found. The aim of this work was precisely the same as ours, to use these intermediate-age clusters to set a lower bound on the ages of RRL stars. Having found no RRLs in the younger clusters, they concluded `that RR Lyraes do not occur in SMC clusters younger than  $\sim$11 Gyr'. They were motivated to probe this particular question by results from \citet{Strugnell1986} who found a population of solar neighbourhood RRL stars to have higher metallicities and angular momenta (around the Galactic centre, i.e. systemic rotation velocities) similar to thin disc stars, which they suggest may be indicative of younger ages than previously thought for these stars, following a very similar line of reasoning as I21,  
and indeed ours, over 35~years ago.   

\section{Conclusions}\label{Sec:conclusions}

In this work we have revisited the possible existence of RRL stars at intermediate ages (1 to 8 Gyr) by searching for RRLs associated to intermediate age clusters in the LMC and SMC. Using the Gaia DR3 SOS \citep{Clementini2023_SOS_GaiaDR3} and OGLE-IV \citep{Soszynski2016} catalogues of RRLs and the database of Structural Parameters of Local Group Star Clusters for LMC and SMC clusters\footnote{\url{https://people.smp.uq.edu.au/HolgerBaumgardt/globular/lgclusters/parameter.html}}, we obtained membership probabilities for RRL stars around each cluster based on proper motion,  distance and angular separation from the cluster's centre (Sec.~\ref{Sec:model}).  We identified \NRRintProb\ RRL stars as probable members ($p>0.5$ and $\Delta<3R_h$) of 10 intermediate age clusters: 3 in the SMC and 7 in the LMC. Information for these RRL, along with that for the probable members of old clusters, is listed in Table~\ref{Tab:cands}.

Using the expectation value for the number of RRLs obtained for each cluster, along with the cluster's initial mass, we used the probabilistic model described in Sec.~\ref{Sec:dtd_inference_model} to infer the \neww{decontaminated} DTD for cluster RRLs (i.e. the number of RRLs per unit initial mass expected a certain time after a burst of star formation) in three age ranges: 1.3 to 2 Gyr (young), 2 to 8 Gyr (intermediate) and >8 Gyr (old). For the old clusters we found $2.5^{+0.4}_{-0.3}$ RRL$/10^5 M_\odot$, in agreement with the DTD inferred for the old field population of the LMC by S21. For the young and intermediate age clusters we found $0.34^{+0.17}_{-0.12}$ and $0.071^{+0.073}_{-0.041}$ RRL$/10^5 M_\odot$, respectively. In both cases these values are lower than those reported by S21 for the field LMC population in the respective age bins, \neww{and should be taken as lower bounds on the DTD as contamination has been accounted for, but incompleteness of the RRL catalogues has not}.

We have also provided the inferred PTF or present time frequency of RRL, i.e. the number of RRLs per \emph{present-day} stellar mass. This should prove useful when estimating the expected number of young or intermediate age RRL stars expected in other populations such as stellar clusters or dwarf galaxies, for which current stellar mass is often reported. These results, along with those for the DTD are summarised in Table~\ref{tab:DTD_PTF} and shown in Figure~\ref{fig:DTD_PTF}.

Final confirmation of the intermediate-age nature of the RRL reported here awaits spectroscopic follow-up to obtain radial velocities \neww{and metallicities} that will robustly confirm membership of these stars to the respective intermediate age clusters. This is a challenge given the pulsating nature of the stars and how distant ($\sim$50 and 60~kpc) and faint ($G\sim19-20$) they are, but an interesting one as it may provide the first direct proof of the existence of these puzzling hypothetical RRL stars. 

\section*{Acknowledgements}

 The authors are happy to thank the anonymous referee for a critical review and constructive exchange that contributed to enhance the presentation of the present work. CM and BCO are glad to thank Sumit Sarbadhicary, Carles Badenes, Bruno Dom\'inguez, Mauro Cabrera-Gadea and Vasily Belokurov for useful discussions. CM is also particularly grateful to Adrian Price-Whelan for the revelation of \verb|numpyro|. BCO thanks Secretaría de Ciencia, Humanidades, Tecnología e Innovación (SECIHTI) and Benemérita Universidad Autónoma de Puebla (BUAP) for the support given during this work. BCO also thanks Universidad Nacional Autónoma de México (UNAM) through the DGAPA fellowship program which partially supported this project \newww{and through grant DGAPA/PAPIIT IN106124}. GB acknowledges financial support from UNAM through grants DGAPA/PAPIIT \newww{IG100319, BG100622 and IN106124}. This research has been supported by funding from project FCE\_1\_2021\_1\_167524 of the Fondo Clemente Estable, Agencia Nacional de Innovaci\'on e Investigaci\'on (ANII). BCO acknowledges funding from the MIA program at Universidad de la Rep\'ublica (UdelaR), Uruguay, and is grateful for the hospitality and support of the Instituto de F\'isica, Facultad de Ciencias, UdelaR where part of this research was carried out. This study was supported by the Klaus Tschira Foundation. The authors thankfully acknowledge the computer resources, technical expertise and support provided by the Laboratorio Nacional de Supercómputo del Sureste de México, CONACYT member of the network of national laboratories. This work has made use of data from the European Space Agency (ESA) mission {\it Gaia} (\url{https://www.cosmos.esa.int/gaia}), processed by the {\it Gaia} Data Processing and Analysis Consortium (DPAC, \url{https://www.cosmos.esa.int/web/gaia/dpac/consortium}). Funding for the DPAC has been provided by national institutions, in particular the institutions participating in the {\it Gaia} Multilateral Agreement.


\section*{Software}
    astropy \citep{astropy2018},
    matplotlib \citep{mpl},
    numpy \citep{numpy},
    scipy \citep{scipy2001},
    jupyter \citep{jupyter2016}, 
    emcee \citep{ForemanMackey2013_emcee_ASCL,ForemanMackey2013_emcee_PASP},
    numpyro \citep{numpyro_bingham2019,numpyro_phan2019},
    imf (\url{https://github.com/keflavich/imf})
    and 
    TOPCAT \citep{Topcat2005,Stilts2006}

\section*{Data Availability}

The input catalogues of RR Lyrae stars and stellar clusters in the LMC and SMC used in this research are based on publicly available catalogues, as described in Sec.~\ref{Sec:Data}. The list of RRL members likely associated to each stellar cluster is made available in Table~\ref{Tab:cands}.



\bibliographystyle{mnras}
\bibliography{refsdtd,refssoftware} 


\appendix

\section{Proper motions and positions of RRL member candidates}\label{Ap:pm_and_pos}

\begin{figure*}
    \centering
    \includegraphics[width=\textwidth]{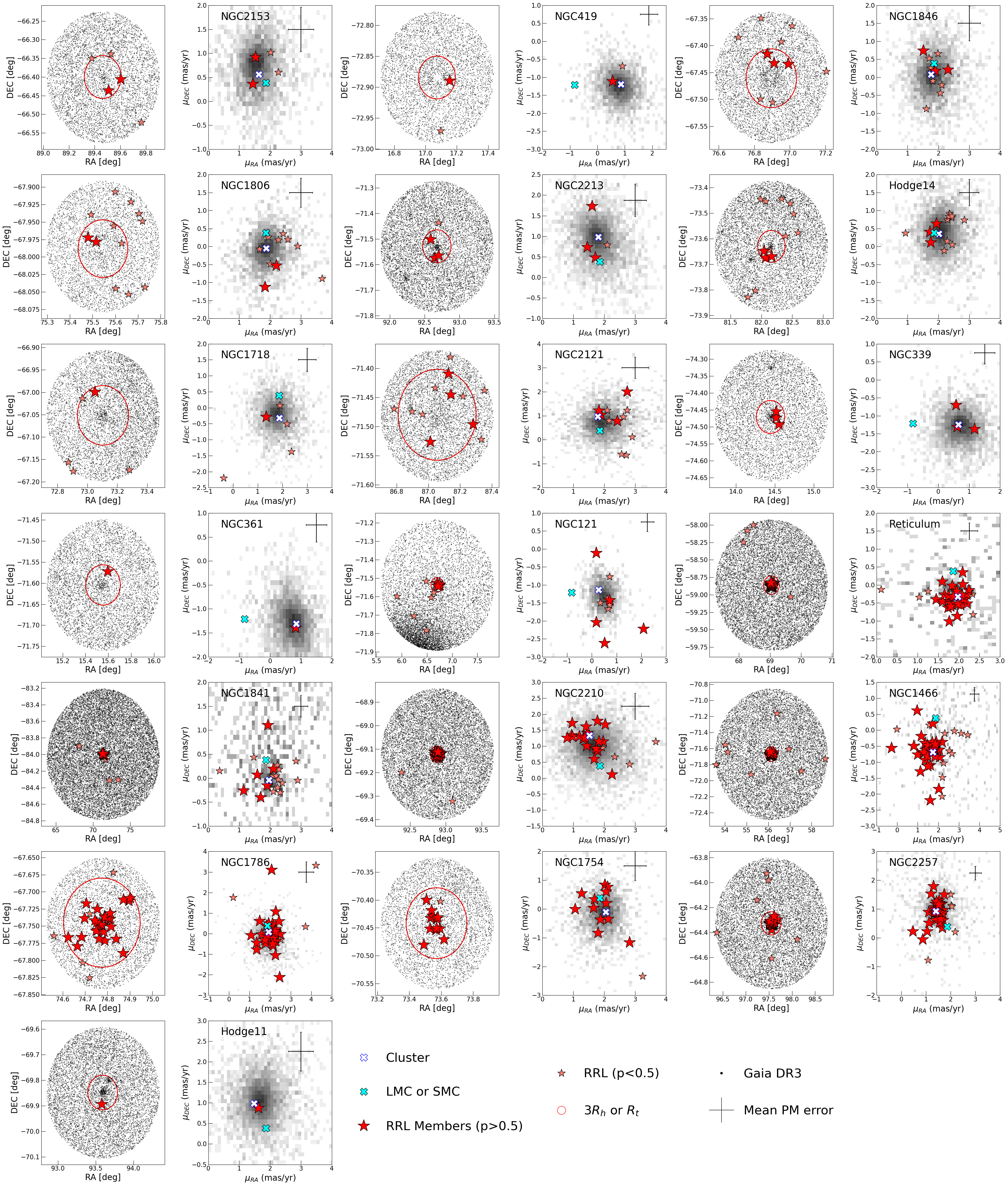}
    \caption{Positions and proper motions of the RRL (low and high-probability) plotted over field stars from Gaia DR3. A red circle is shown for reference purposes, indicating the outer boundary of the cluster (considered to be three times the half-mass radius ($R_h$) for the majority of clusters, and for clusters with small $R_h$ values, the circle indicates the tidal radius in the literature). The member RRLs in red correspond to the stars in Table~\ref{Tab:cands}. }
    \label{fig:ap_pm_and_pos}
\end{figure*}

Figure \ref{fig:ap_pm_and_pos} shows celestial positions and proper motion distributions of the RRL member candidates for clusters with at least one high membership probability star.

\section{Light curves}\label{Ap:lcs}

\begin{figure*}
    \centering
    \includegraphics[width=1.7\columnwidth]{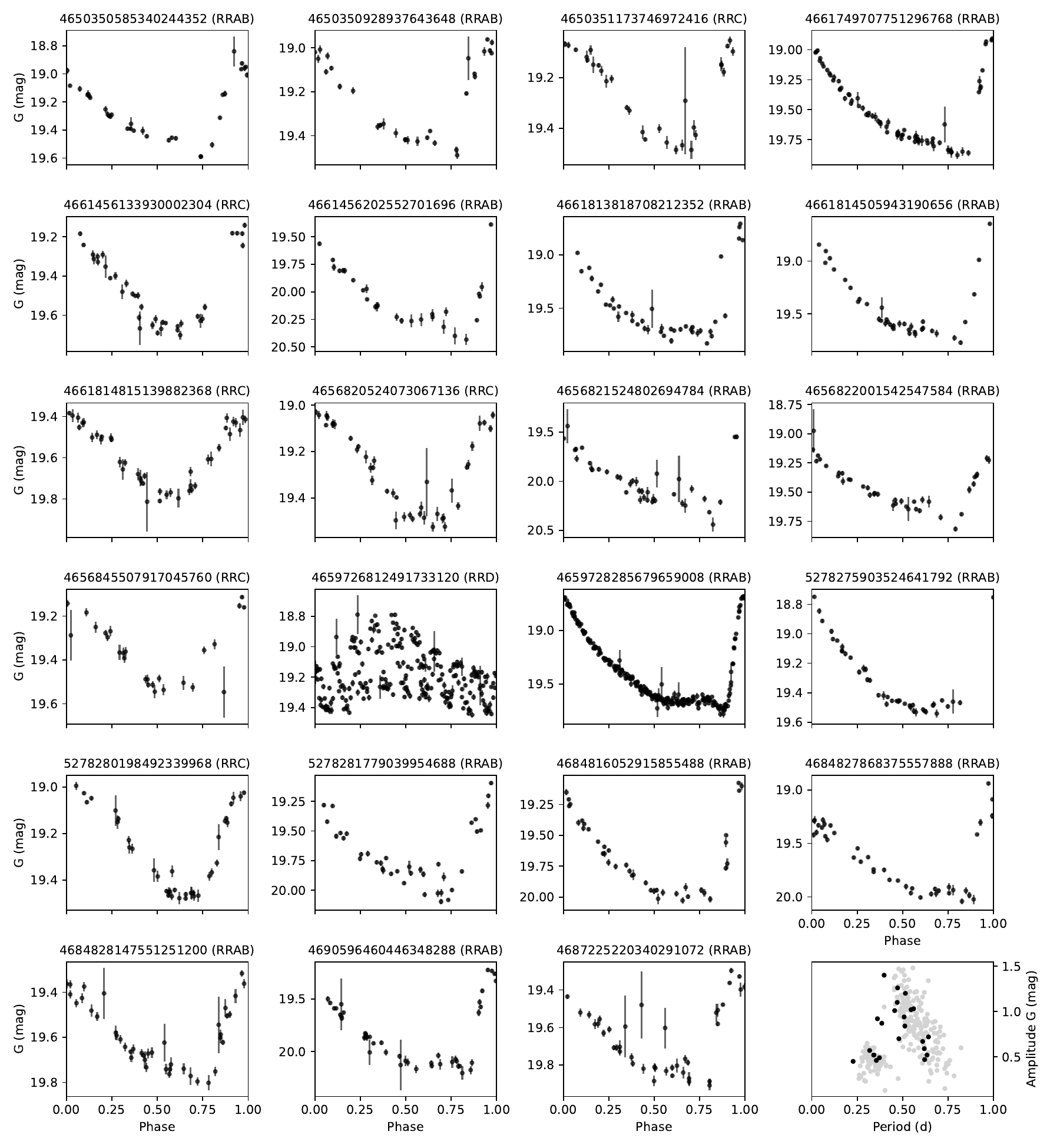}
    \caption{Phase-folded G-band light curves for the \NRRintProb\ RRL stars from Table~\ref{Tab:cands} with high probability ($p>0.5$) of membership to clusters with intermediate ages (1 to 8 Gyr). The last panel (bottom right) shows the Period-Amplitude (G-band) diagram for these \NRRintProb\ stars (black) in comparison to the rest of cluster member candidate RRLs (gray). }
    \label{fig:lcs}
\end{figure*}

Figure~\ref{fig:lcs} shows G-band light curves for the \NRRintProb\ RRL stars with high probability ($p>0.5$) of membership to clusters with intermediate ages (1 to 8 Gyr) reported in Table~\ref{Tab:cands}.

\bsp	
\label{lastpage}
\end{document}